\documentclass{aa}
\usepackage{graphicx}
\usepackage{txfonts}
\usepackage{natbib}
\begin{document}

   \title{Interferometric observations of the Mira star $o$~Ceti with the VLTI/VINCI instrument in the near-infrared
\thanks{Based on observations collected at the European Southern Observatory, Paranal, Chile (public commissioning data)}\fnmsep
\thanks{Based on data collected at the Special Astrophysical Observatory (SAO), Russia}\\}

\author{H.~C.~Woodruff\inst1,
 M.~Eberhardt\inst1,
 T.\,Driebe\inst1,
 K.-H.\,Hofmann\inst1,
 K.\,Ohnaka\inst1,
 A.\,Richichi\inst2,
 D.\,Schertl\inst1,
 M.\,Sch\"oller\inst3,
 M.\,Scholz\inst4\inst,\inst5,
 G.\,Weigelt\inst1,
 M.\,Wittkowski\inst2, and
P.~R.\,Wood\inst6}

   \institute{
     \inst1 Max-Planck-Institut f\"ur Radioastronomie, Auf dem H\"ugel 69, D-53121 Bonn, Germany\\
     \inst2 European Southern Observatory, Karl-Schwarzschildt-Str. 2, D-85748 Garching, Germany\\ 
     \inst3 European Southern Observatory, Casilla 19001, Santiago 19, Chile\\
     \inst4 Institut f\"ur Theoretische Astrophysik, Universit\"at Heidelberg, Tiergartenstr.~15, D-69121 Heidelberg, Germany\\
     \inst5 Institute of Astronomy, School of Physics, University of Sydney, NSW 2006, Australia\\
     \inst6 Research School of Astronomy and Astrophysics, Australian National University, Cotter Road, Weston Creek ACT 2611, Australia\\
         }

\offprints{ H.~C.~Woodruff \\ \email{woodruff@mpifr-bonn.mpg.de}}

  \date{Received ; accepted }

\authorrunning{H.C.\ Woodruff et al.}
\titlerunning{VINCI observations of Mira}

\abstract{
We present \textit{K}-band commissioning observations of the Mira star prototype $o$ Cet 
obtained at the ESO Very Large Telescope Interferometer (VLTI) 
with the VINCI instrument and two siderostats. 
The observations were carried out between 2001 October and December, in 2002 January and December, 
and in 2003 January. Rosseland angular radii are derived from the measured visibilities 
by fitting theoretical visibility functions obtained from center-to-limb intensity variations 
(CLVs) of Mira star models \citep{BSW,HSW,TLSW}.
Using the derived Rosseland angular radii and the SEDs reconstructed from available photometric 
and spectrophotometric data, we find effective temperatures
ranging from $T_{\rm eff}=3192 \pm 200$ K at phase $\Phi=0.13$ to $2918 \pm 183$ K at $\Phi=0.26$.
Comparison of these Rosseland radii, effective temperatures, and the shape of the observed 
visibility functions with model predictions suggests that $o$ Cet is a fundamental mode pulsator.
Furthermore, we investigated the variation of visibility function and diameter with phase. 
The Rosseland angular diameter of $o$~Cet increased from $28.9 \pm 0.3$~mas
(corresponding to a Rosseland radius of $332 \pm 38 ~R_{\odot}$ for a distance of $D=107\pm12~{\rm pc}$) 
at $\Phi=0.13$ to $34.9 \pm 0.4$~mas ($402 \pm 46~R_{\odot}$) at $\Phi=0.4$. 
The error of the Rosseland linear radius almost entirely results from the error of the parallax, 
since the error of the angular diameter is only approximately 1\%.
\keywords{
instrumentation: interferometers --
techniques: interferometric -- 
stars: late-type  -- 
stars: AGB and post-AGB --
stars: fundamental parameters --
stars: individual: Mira}
}

   \maketitle

   \section{Introduction}\label{intro}
%
Mira stars are long-period variables (LPVs) which evolve along the
asymptotic giant branch (AGB), and are characterized by stellar pulsation 
with amplitudes as large as $\Delta V \sim 9$ and well-defined pulsation
periods (80-1000 days).  
In recent years, the comparison of theoretical pulsation models with MACHO 
observations of LPVs in the LMC, in particular the reproduction 
of period ratios in multimode
pulsators, has shown that Miras are fundamental-mode pulsators
\citep{woodM}.
However, radius measurements of Mira variables when compared to theoretical
pulsation calculations have generally yielded the large values expected for
first overtone pulsators (e.g., \citealt{Feast,VANB}). There is clearly
a problem with the interpretation of radius measurements that needs
examination.\\
High-resolution interferometric studies of Mira stars allow the determination
of the size of the stellar disk, its center-to-limb intensity
variation, surface inhomogeneities, and the dependence of diameter
on wavelength and variability phase
(see, e.g., \citealt{PEA}; \citealt{BON}; \citealt{LAB77};
 \citealt{BON82}; \citealt{KAR}; \citealt{HAN92}; \citealt{QUI}; \citealt{WIL92}; \citealt{TUT94};
 \citealt{DAN}; \citealt{HAN95}; \citealt{WEI96}; \citealt{VANB}; \citealt{BURNS}; \citealt{PER}; 
\citealt{HOF00}; \citealt{WEI00}; \citealt{WEIN}; \citealt{THOM}; \citealt{MEIS}).
The results of such interferometric observations can be compared 
with predictions from theo\-retical models of 
stellar pulsation and the atmosphere of 
Mira stars (e.g., 
\citealt{WAN}; \citealt{SCHO}; \citealt{BES}; 
\citealt{BSW} = BSW;  \citealt{HSW} = HSW; \citealt{TLSW} = TLSW;
\citealt{ISW} = ISW).  
Confrontation of detailed theoretical models with
high-resolution observations 
is crucial for improving our understanding of the physical 
properties of Mira stars (e.g., \citealt{HOFRCAS}; \citealt{HOF01}; \citealt{WEI03}; \citealt{SCH03}). \\
In this paper we present ESO VLTI/VINCI visibility measurements of $o$~Cet 
and compare the measured visibility shape and the phase
dependence of the visibility with model predictions. $o$~Cet, the
prototype of oxygen-rich Mira stars, is a very suitable object
for these studies, since VINCI observations exist
for different phases and baselines, its distance is known (revised
HIPPARCOS distance $107.06\pm12.26~{\rm pc}$, \citealt{KNAPP}), and a large amount of
spectroscopic and photometric data is available for different
phases.
   \section{Observations and data reduction}\label{obsdr}
\subsection{Observations}\label{obs}
A total of 48 visibility measurements of $o$ Cet were carried out with VINCI (\citealt{KER}) 
at the VLTI in the commissioning period between 2001 October and 2003 January.
All this data has been publicly released. Projected baselines ranging from 7.5 to 16.0 m were employed.
VINCI is a fiber-optics beam combiner instrument based on the concept of the FLUOR 
instrument \citep{CR,MAR}. 
With the single-mode fibers to spatially 
filter the wavefronts perturbed by atmospheric turbulence, 
this beam combiner provides accurate visibility measurements in 
spite of time-variable atmosphere conditions. 

\begin{figure}[!ht]
\resizebox{\hsize}{!}{\includegraphics[angle=-90]{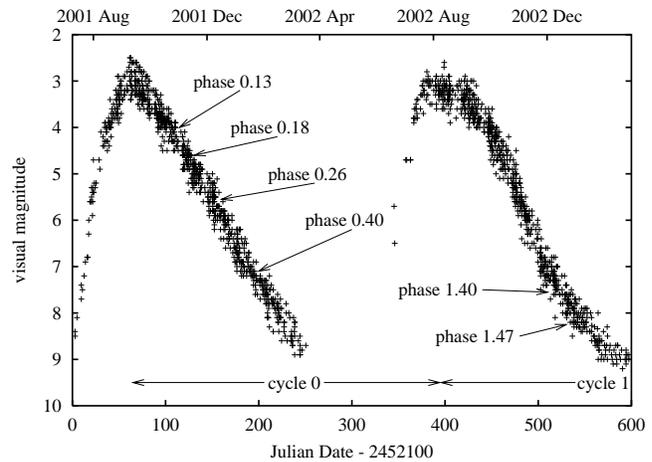}}
\caption{Visual light curve for $o$ Cet \citep{AAVSO} between JD 2452100 and 2452700. 
The VINCI visibilities were measured at the following phases: 
$\Phi = 0.13$ (2001 Oct 22/23), $\Phi = 0.18$ (2001 Nov 09/15/17), 
$\Phi = 0.26$ (2001 Dec 06), $\Phi = 0.40$ (2002 Jan 20), $\Phi = 1.40$ (2002 Dec 20), 
and $\Phi = 1.47$ (2003 Jan 09).
Phases larger than 1 indicate that the observations belong to the next pulsation cycle.}
\label{lightcurve}
\end{figure}

Table \ref{obs1} gives an overview of the VINCI observations. 
Figure \ref{lightcurve} shows the visual light curve of $o$ Cet together with the dates at which the 
VINCI observations were carried out \citep{AAVSO}. 
Figure~\ref{vis} shows the visibilities of $o$~Cet vs. spatial frequency measured at six different phases. 
The comparison between the observations and the different Mira star model series will be discussed 
in Sect.~\ref{mods}. 

In addition to the VINCI observations, we recorded
speckle interferograms of $o$ Cet on 2003 October 7 with the SAO 6\,m telescope in Russia.
The speckle camera used for the observations was equipped with a Rockwell HAWAII array.
The field of view of the recorded speckle interferograms was $11\farcs02$. 
A filter with a central wavelength of 2086\,nm and a width of 20\,nm was used. 
The exposure time per frame was 10 ms.
The complete data set consists of 100 speckle interferograms of $o$ Cet and 360 of an 
unresolved reference star (HD 14652). The seeing (FWHM) was $\sim 3\farcs4$. 
The visibilities were obtained using the speckle interferometry method (\citealt{LAB70}).
\begin{table}[!ht]
\begin{center}
\caption {Summary of VINCI commissioning observations of $o$ Cet: 
Date, Julian Date JD, cycle and visual phase $\Phi$ 
(see Fig.~\ref{lightcurve}), number of visibility data points $N$, projected baseline 
length $B_{\rm p}$, and baseline projection angle P.A.}
\label{obs1}
\begin{tabular}{l c c c c c}\hline
Date & JD & cycle+$\Phi$ & $N$ & $B_{\rm p}$ [m] & P.A. [\degr]\\ \hline
2001 Oct 22 & 2452205 & 0.13 & 13 & 11.5 - 16.0 & 62-73\\ 
2001 Oct 23 & 2452206 & 0.13 & 12 & 11.5 - 16.0 & 63-73\\ 
2001 Nov 09 & 2452223 & 0.18 & 6  & 14.5 - 16.0 & 71-73\\ 
2001 Nov 15 & 2452229 & 0.18 & 3  & 15.5 - 16.0 & 72-73\\
2001 Nov 17 & 2452231 & 0.18 & 3  & 14.5 - 15.0 & 70\\ 
2001 Dec 06 & 2452250 & 0.26 & 6  & 15.0 - 16.0 & 72-73\\
2002 Jan 20 & 2452295 & 0.40 & 2  & 14.0 - 14.5 & 72\\
2002 Dec 20 & 2452629 & 1.40 & 2  & 7.5   - 8.0 & 73-74\\
2003 Jan 09 & 2452649 & 1.47 & 1  & 8.0         & 73\\ \hline
\end{tabular}
\end{center}
\end{table}
%
\subsection{Data reduction}\label{dr}
\begin{figure*}[!ht]
  \begin{tabular}{p{75mm}l}
\hspace{-1cm} \includegraphics[angle=-90,width=10.7cm]{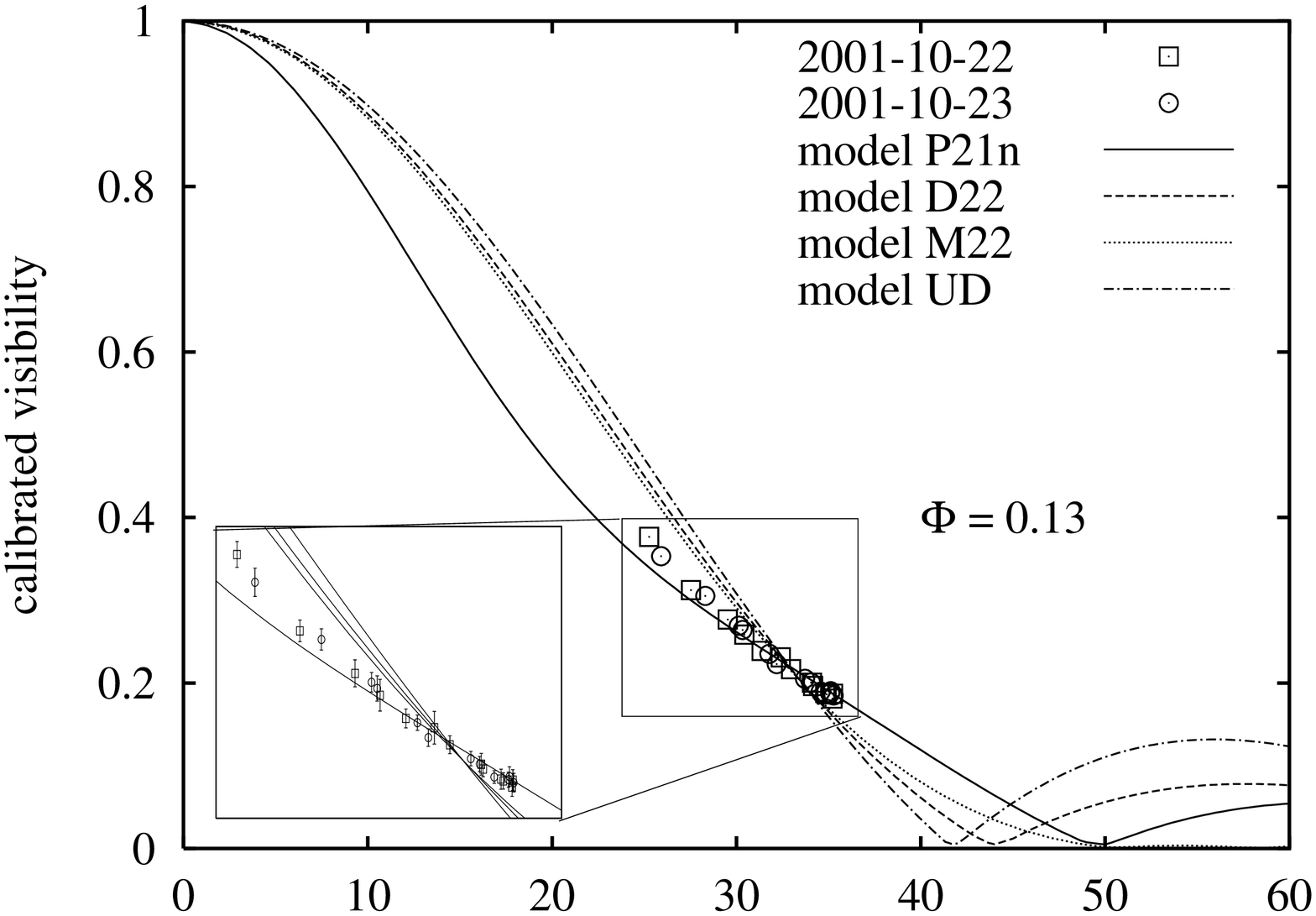}&
 \vspace{-1.5cm} \includegraphics[angle=-90,width=10.7cm]{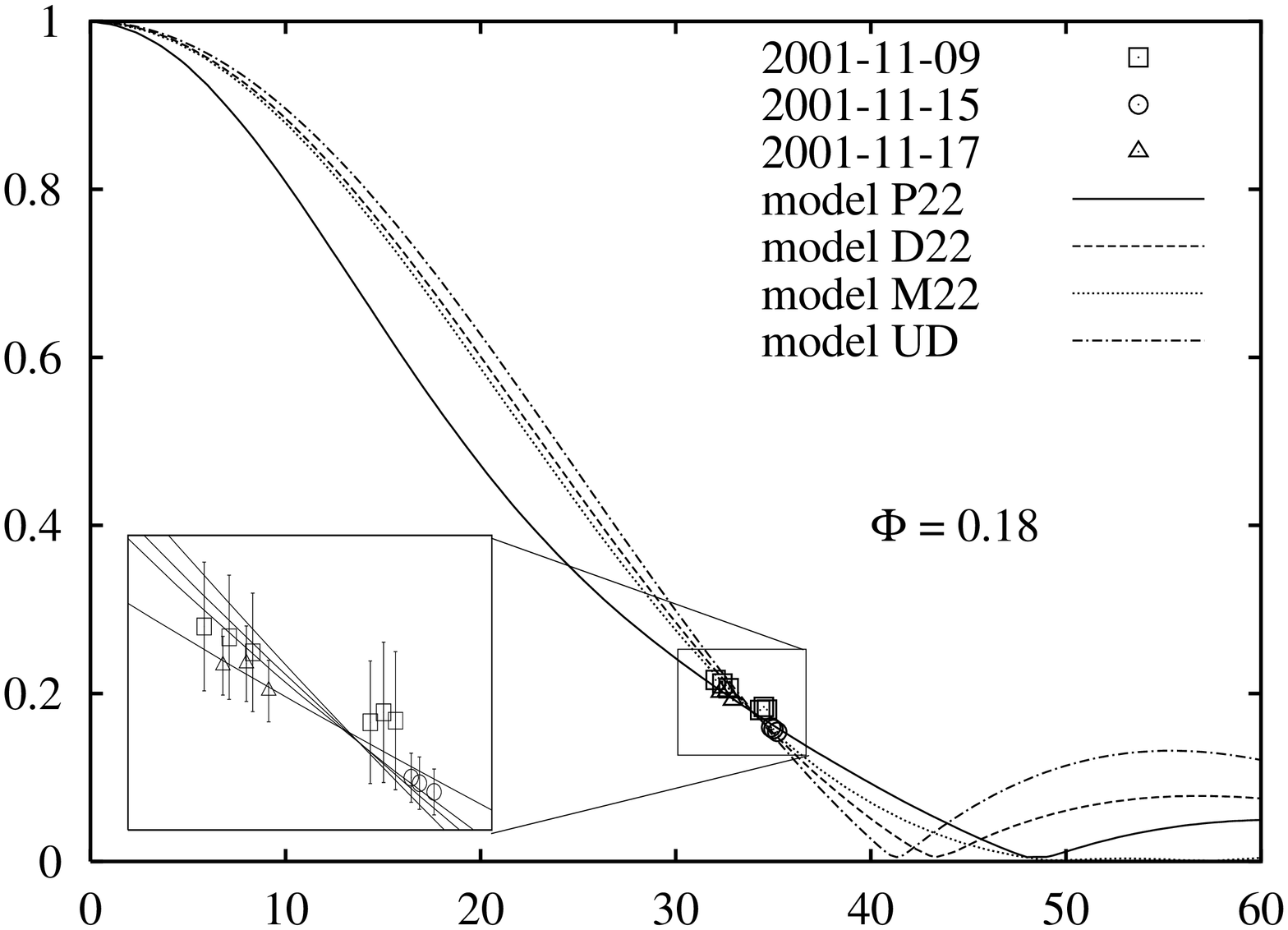}\\
 \hspace{-1cm} \includegraphics[angle=-90,width=10.7cm]{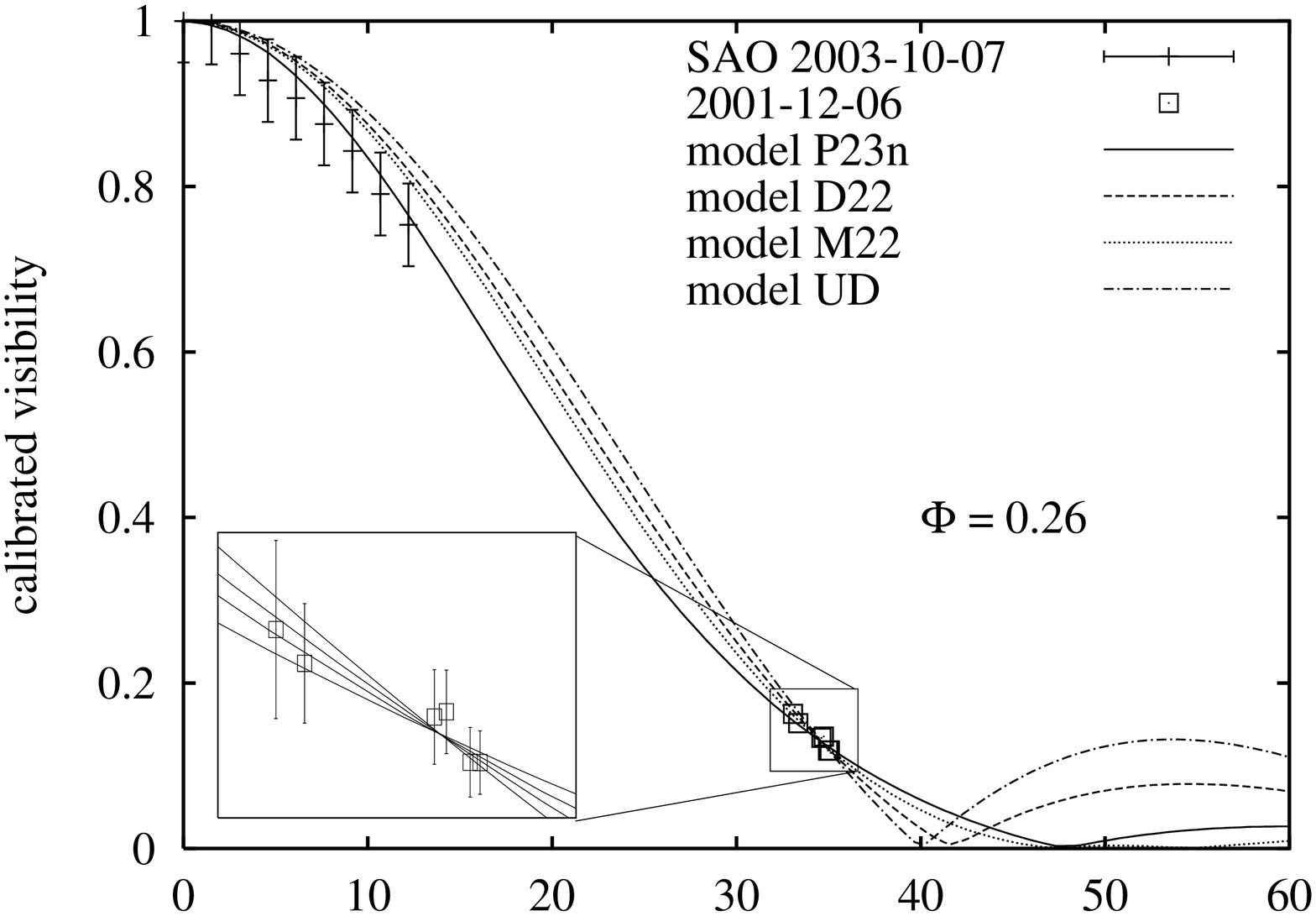}&
 \vspace{-1.5cm} \includegraphics[angle=-90,width=10.7cm]{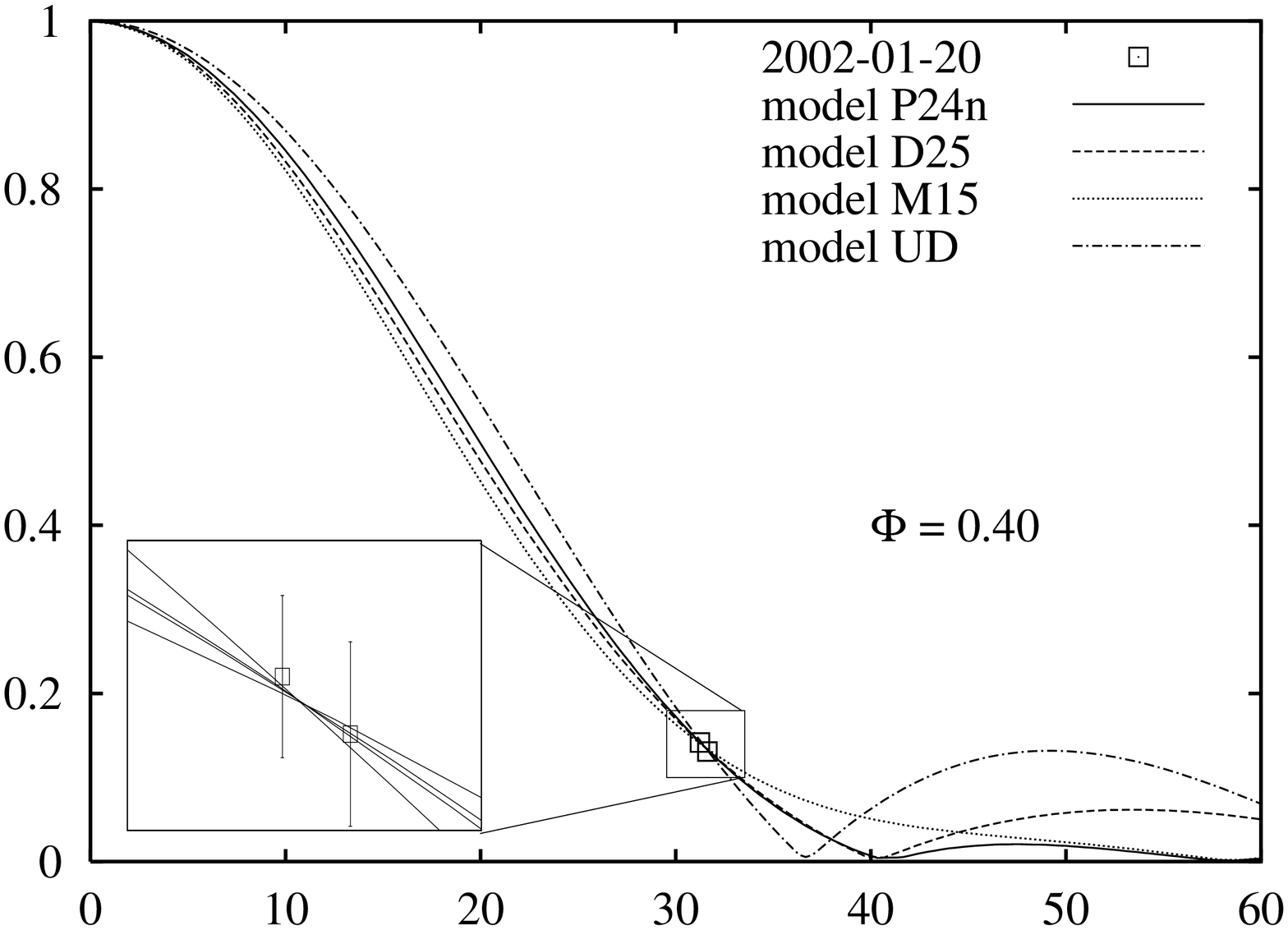}\\
 \hspace{-1cm} \includegraphics[angle=-90,width=10.7cm]{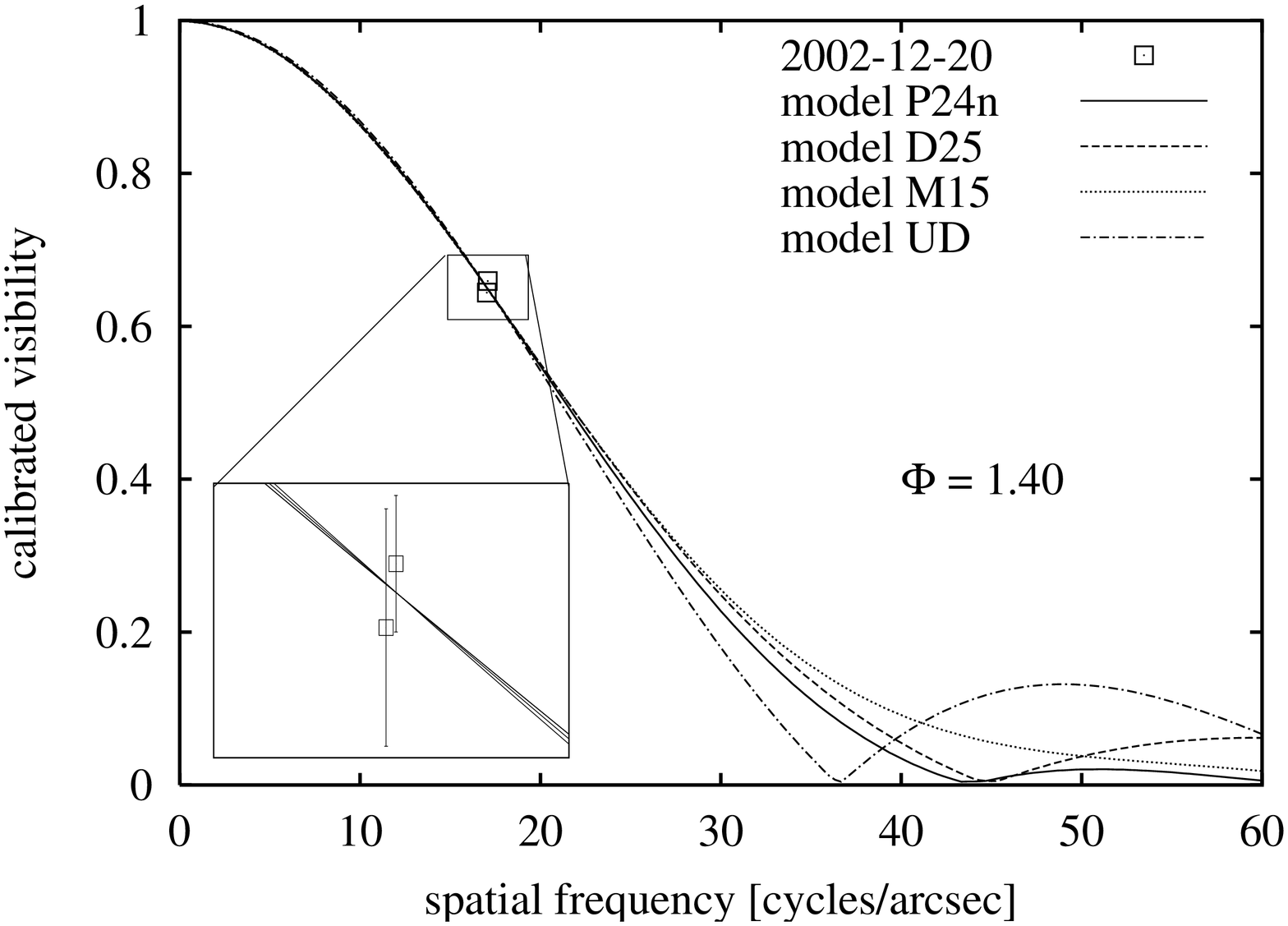}&
 \vspace{-1.5cm} \includegraphics[angle=-90,width=10.7cm]{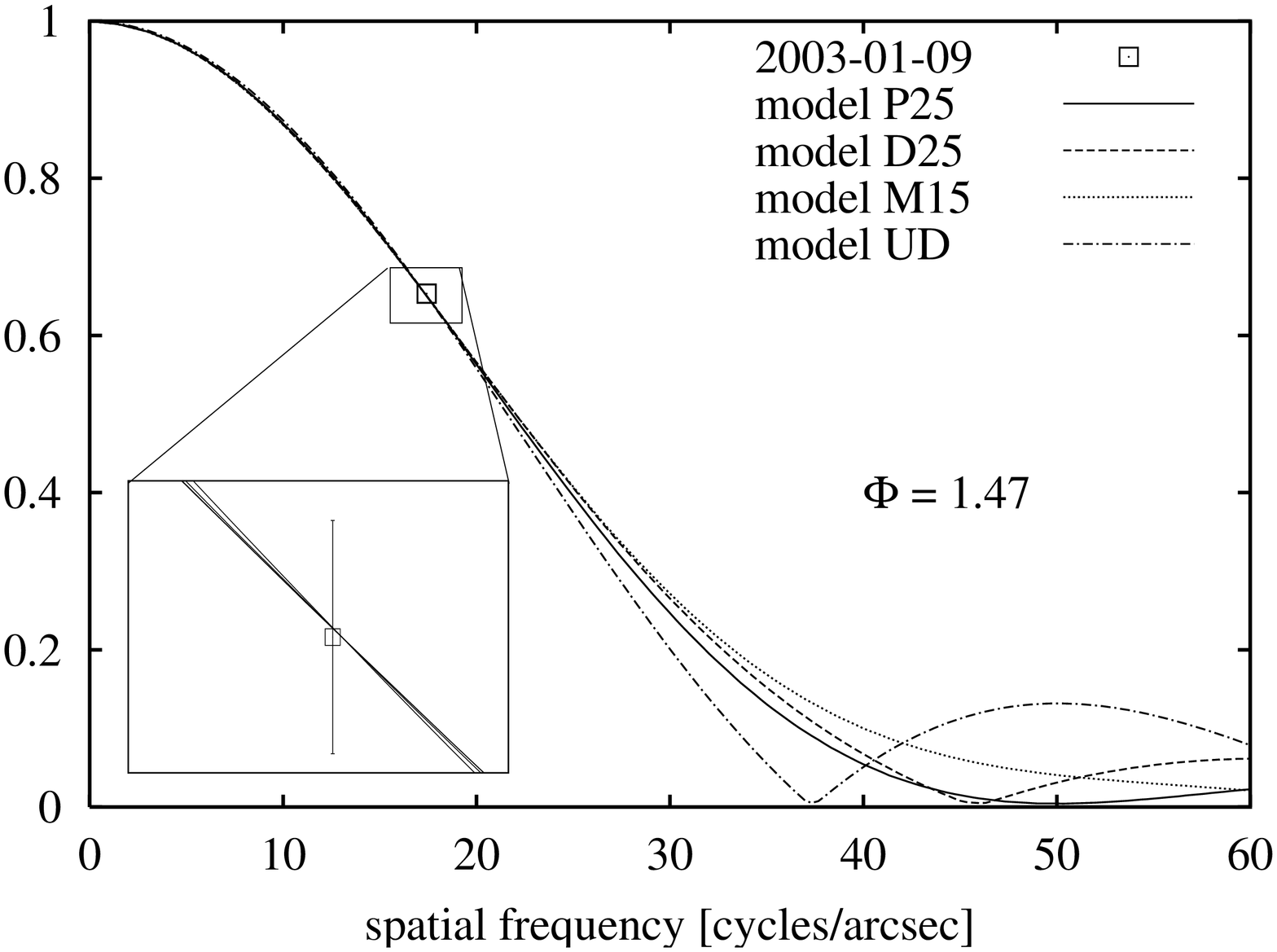}\\
  \end{tabular}
\vspace{0.8cm}
  \caption{Visibilities of $o$~Cet vs. spatial frequency measured at phase 0.13 
({\bf top left}, observation dates 2001 Oct 22/23), at phase 0.18 ({\bf top right}, 
observation dates 2001 Nov 09/15/17), at phase 0.26 ({\bf middle left}, 
observation date 2001 Dec 06) together with SAO speckle interferometry data measured 
at the same phase but at a different cycle 
(observation date 2003 Oct 07; data not used for the model visibility fit), 
at phase 0.40 ({\bf middle right}, observation date 2002 Jan 20), at phase 1.40 
({\bf bottom left}, observation date 2002 Dec 20), and at phase 1.47 
({\bf bottom right}, observation date 2003 Jan 09). 
The insets show an enlargement of the relevant spatial frequency range. Here, the error 
bars are included. The fits with different Mira star models are discussed in Sect.~\ref{mods}. 
Solid lines represent the P models from TLSW and ISW, dashed lines the D models from BSW, 
dotted lines the M models from HSW,
and the dash-dotted lines represent the simple uniform-disk model CLVs.
Note that from our SAO observations there is no indication for a visibility 
contribution from a circumstellar dust shell at low spatial frequencies.}
  \label{vis}
\end{figure*}

\begin{figure*}[!ht]
 \centering
  \begin{tabular}{c @{\hspace{-5mm}} c @{\hspace{-5mm}} c}
    \includegraphics[angle=-90,width=6cm]{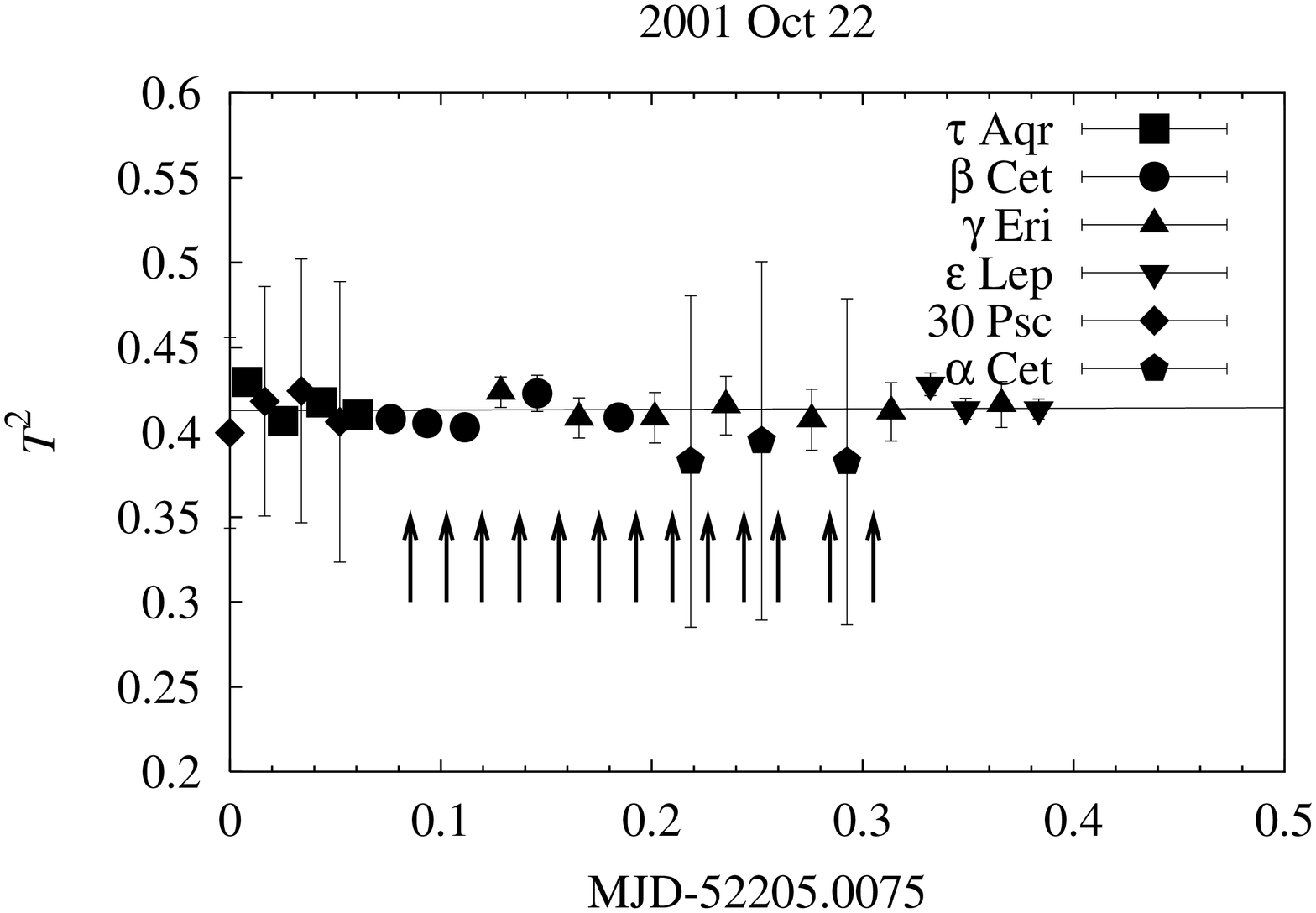}&
    \includegraphics[angle=-90,width=6cm]{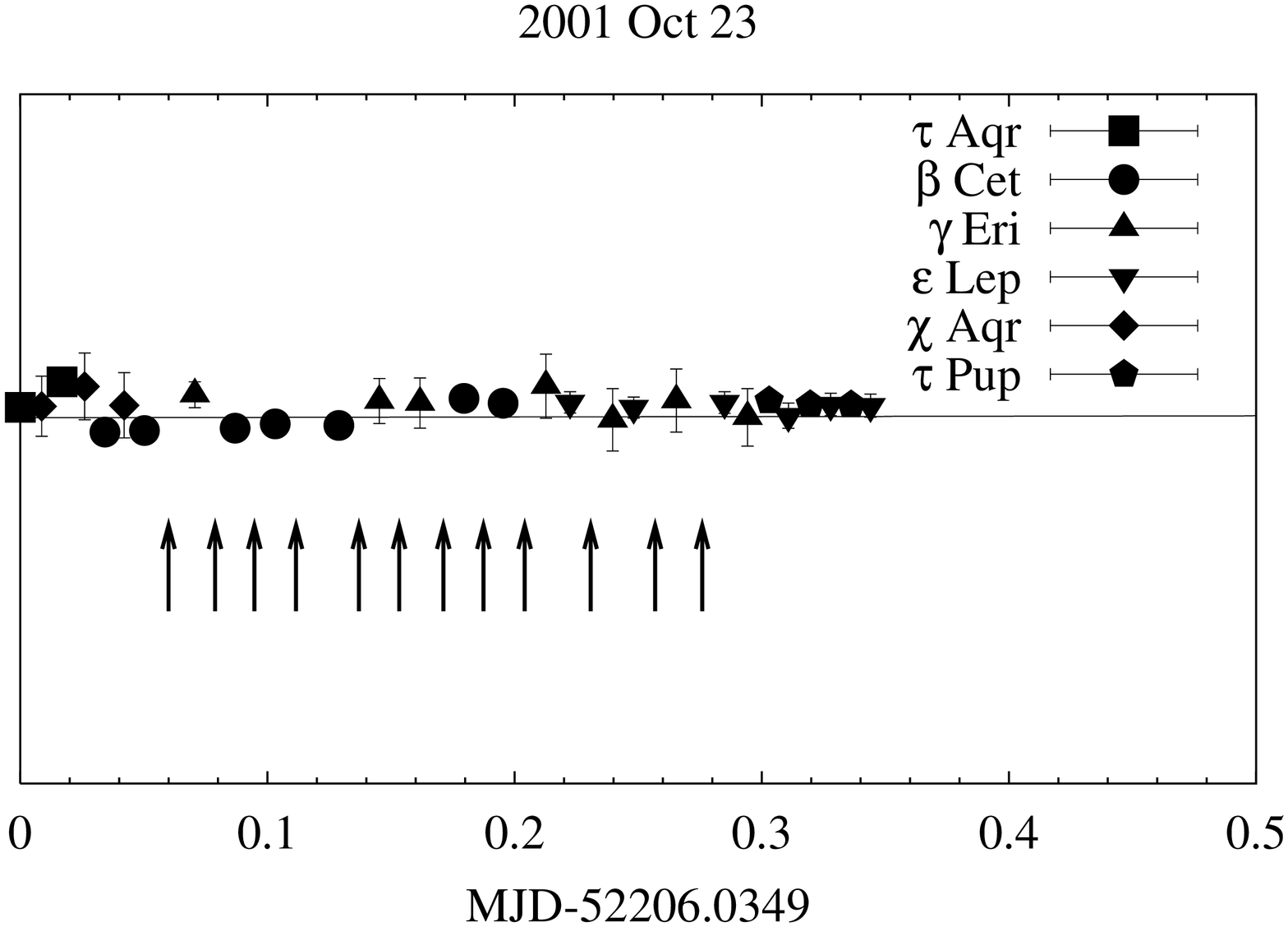}&
    \includegraphics[angle=-90,width=6cm]{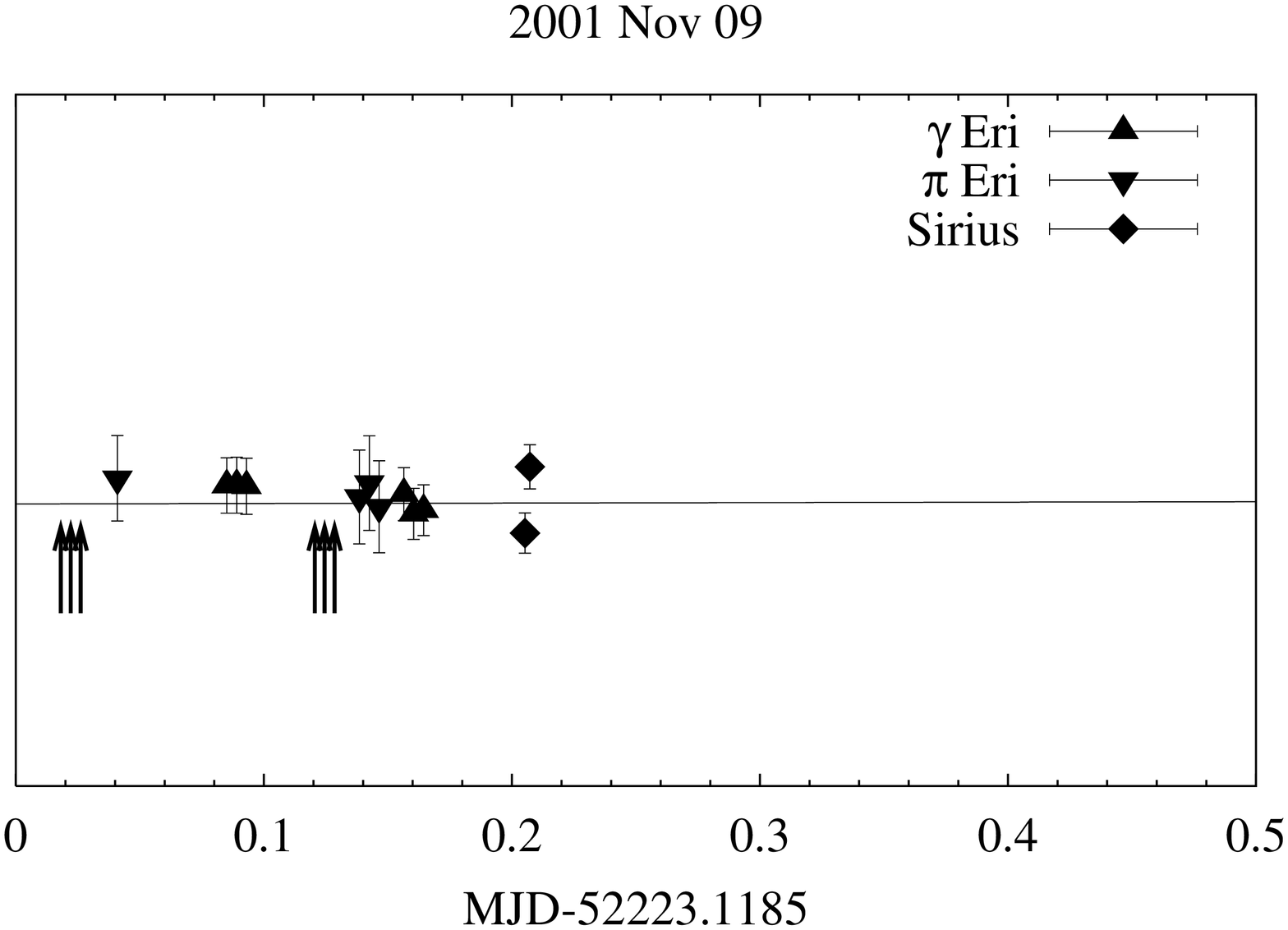}\\
    \includegraphics[angle=-90,width=6cm]{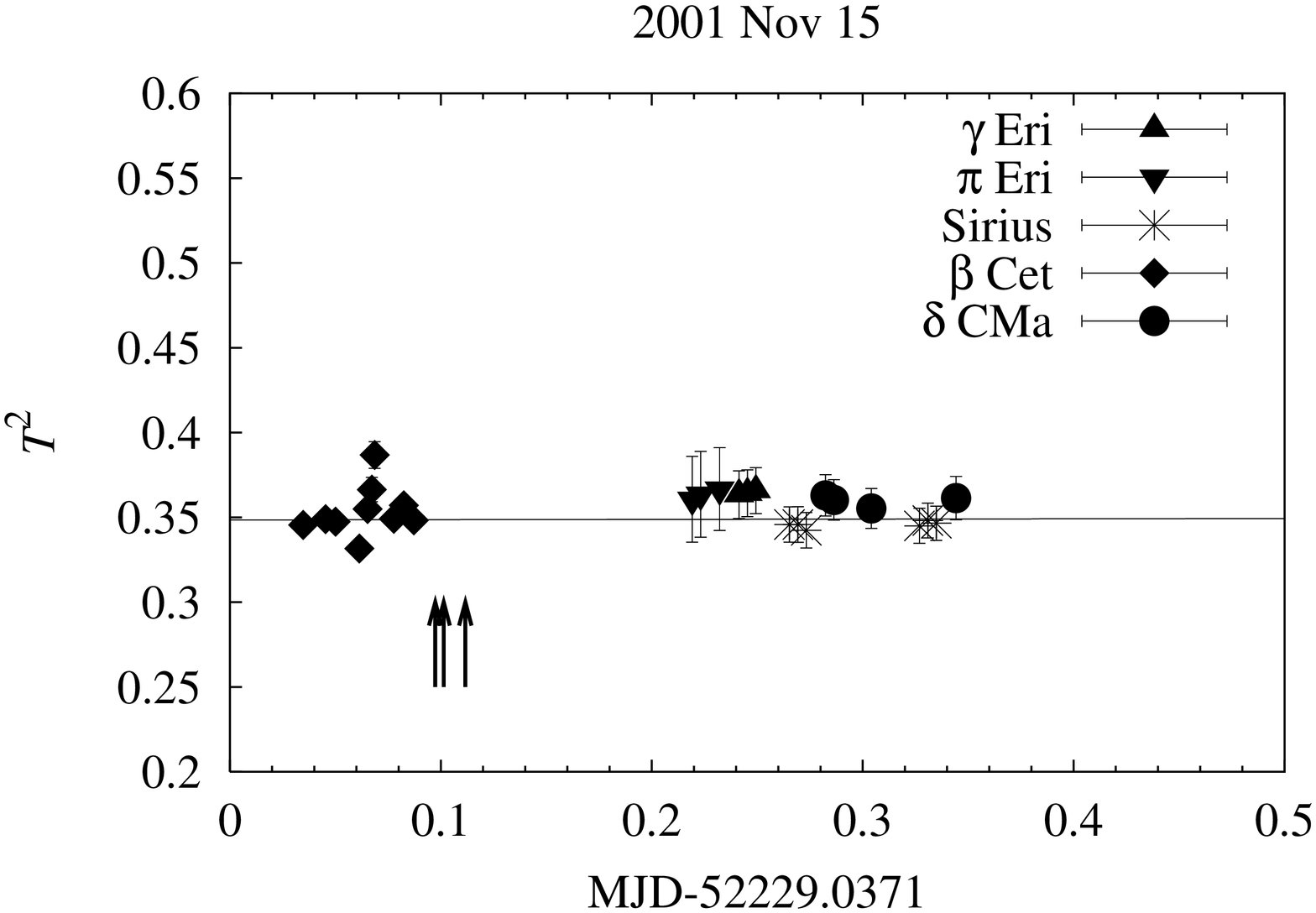}&
    \includegraphics[angle=-90,width=6cm]{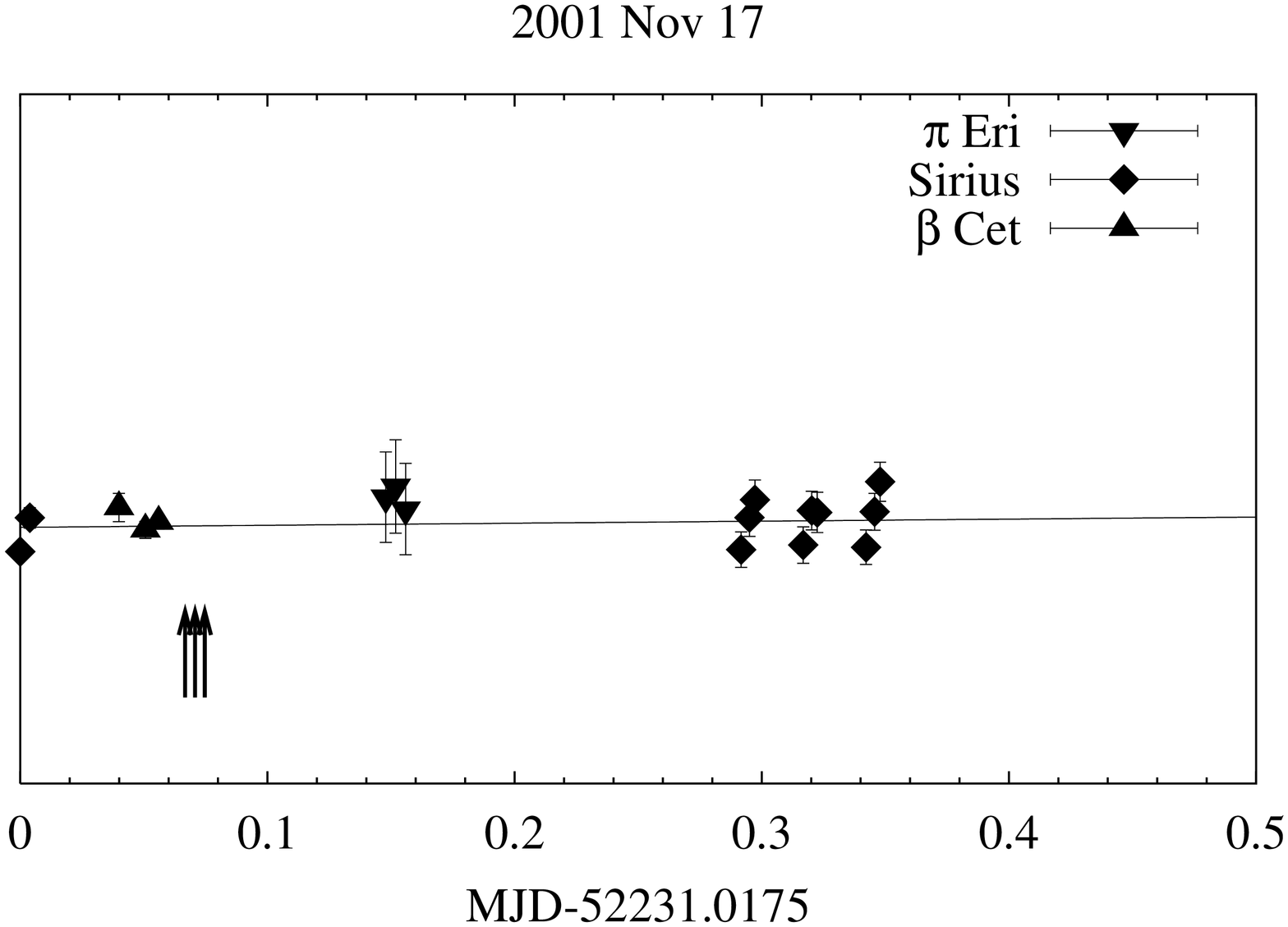}&
    \includegraphics[angle=-90,width=6cm]{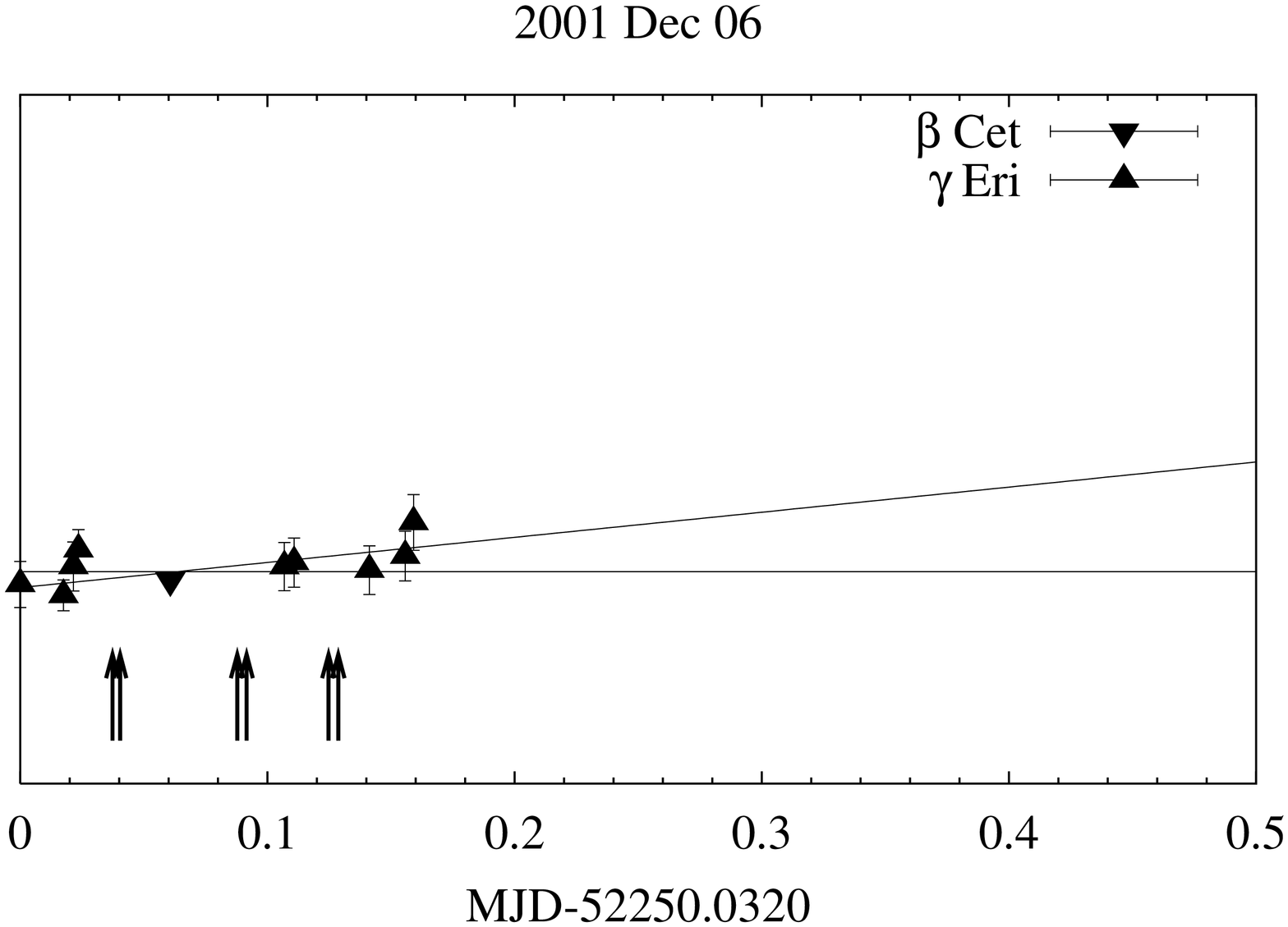}\\
    \includegraphics[angle=-90,width=6cm]{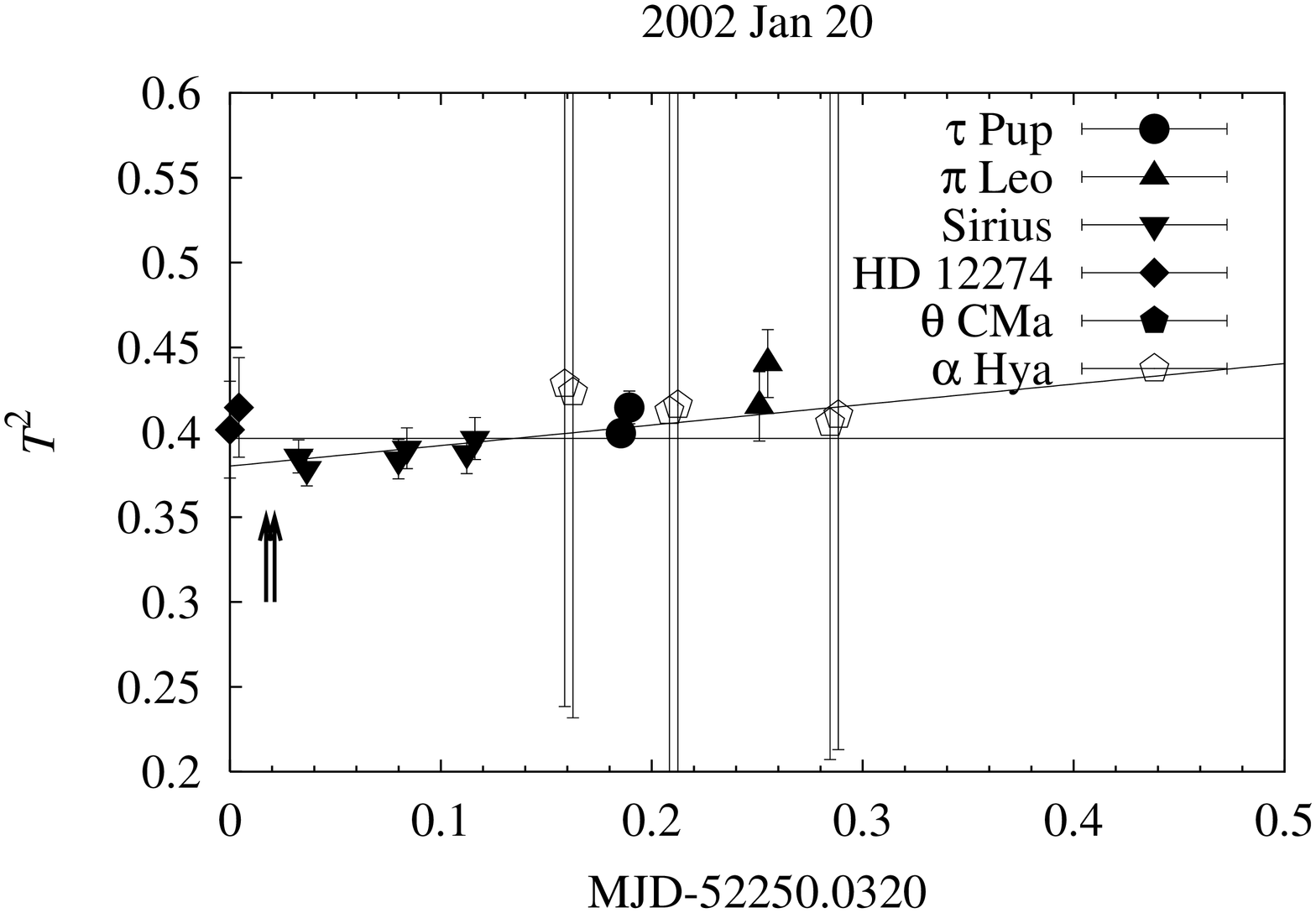}&
    \includegraphics[angle=-90,width=6cm]{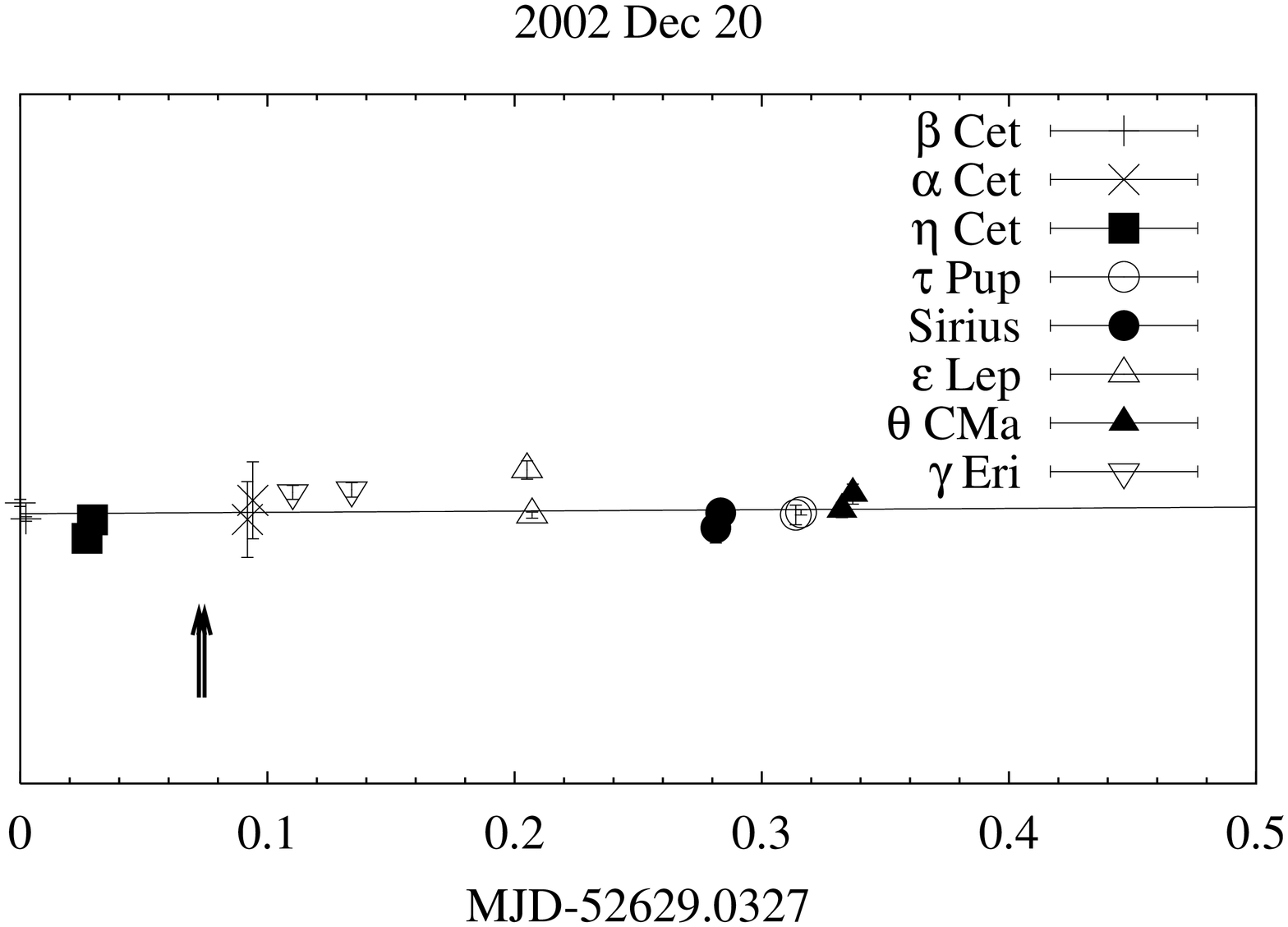}&
    \includegraphics[angle=-90,width=6cm]{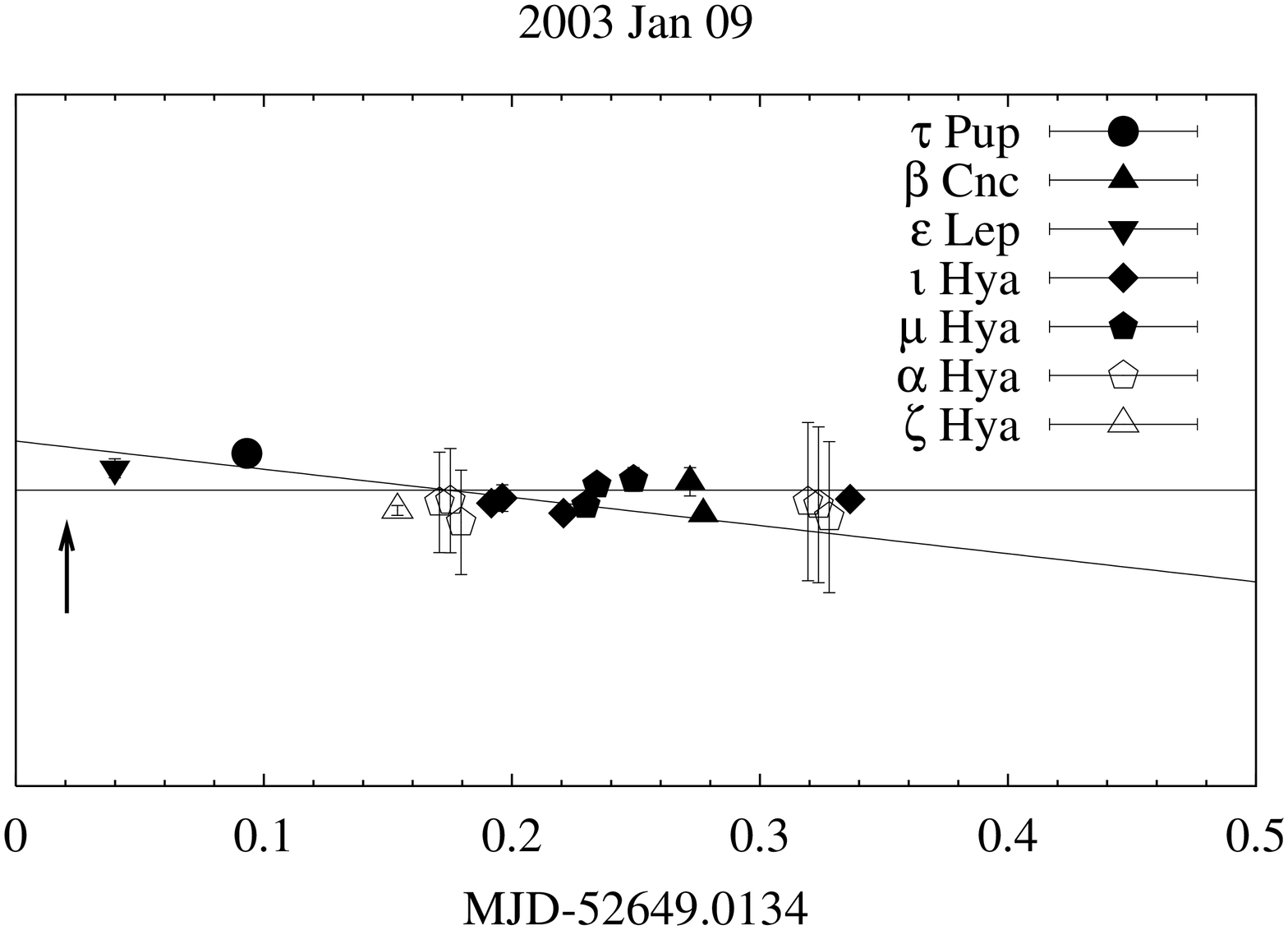}\\
  \end{tabular}
\caption{Linearly fitted squared transfer functions (see text) as a function of time 
(MJD = Modified Julian Date) obtained from calibrator stars measured (see Table~\ref{cal}). 
For comparison, in the nights of 2001 Dec 06, 2002 Jan 20 and 2003 Jan 09 
we included a constant fit of the squared transfer functions. 
The arrows mark the observation times of $o$~Cet.
}
\label{mutf}
\end{figure*}

Processing of the raw VINCI data for estimating the coherence factors $\mu$ 
(i.e.~the fringe contrast or uncalibrated visibility) was carried out 
with the VINCI data reduction software provided by the
European Southern Observatory\footnote{www.eso.org/$\sim$pballest/vinci}
(version 2.0.6), based on wavelets transforms to derive the power spectral density.
The processing of a series of OPD scans yields the uncalibrated squared visibility $\mu^2$ 
for each individual scan, and the averaged value $\langle \mu^2 \rangle$ is used to derive 
the calibrated visibility \citep{FOR,PER2}.\\
The squared transfer functions $T^2$, which account for instrumental and atmospheric effects, 
were derived from measurements of calibrator stars and their known angular diameters.
The expected calibrator visibility $V_{\rm cal}$ was calculated from uniform disk (UD) 
angular diameters of the calibrators given in \cite{CHARM}, and the transfer function 
was evaluated as
\begin{equation}\label{eq1}
T^2 = \frac{\mu_{\rm cal}^2}{V^2_{\rm cal}},
\end{equation}
where $\mu_{\rm cal}$ is the measured uncalibrated visibility of the calibrator.
The calibrators of the nights on which $o$~Cet was measured are listed in Table \ref{cal}.
According to Eq.~(\ref{eq1}), the errors of the individual transfer function values 
were calculated from the errors of the raw visibilities and the errors of the 
uniform-disk visibilities of the calibrators which arise from the uncertainties of the
corresponding uniform-disk diameters. For most of the individual transfer function values, 
the error contributions of the raw visibility and the uniform-disk 
visibility of the calibrator are of the same order,
resulting in total errors of typically $\la$ 5 percent (see Fig.~\ref{mutf}). 
Only in a few cases the error contribution of either the raw visibility or the
uniform-disk visibility of the calibrator is significantly larger, resulting
in much larger errors of the transfer function values.
\\ 
In order to obtain the squared transfer function for the time of the observation of the object, 
the measured values of the squared transfer function $T^2$ of 
each night were plotted against the time of observation and fitted by a straight line. 
Since the transfer function proved to be a slowly varying function which was rapidly sampled, 
as Fig.~\ref{mutf} shows, we did not apply higher order polynomial fits.
The error of the time-dependent transfer function is given by the error of 
the linear interpolation.
To estimate the error introduced by our method of determining 
the transfer function at the time of the observation of $o$ Cet, 
we also applied constant fits to the three nights where 
the variation of the transfer values was largest (2001 Dec 06, 2002 Jan 20, and 2003 Jan 09).
The difference obtained for the $o$~Cet calibrated visibilities 
using the linear and the constant transfer function fit procedures 
is 1.5\% at most, leading to a final error
of 0.1\,mas for the Rosseland diameter of $o$ Ceti. \\
Since VINCI uses a $K$ broad-band filter,
model visibilities calculated with VINCI's filter response function 
would be appropriate for the comparison of models and measurements.
However, those polychromatic VINCI-filter models cannot be calculated since
monochromatic Mira star CLV models are not readily available. Therefore, instead
of using VINCI-filter model CLVs,
we adopted the following approximative scheme:
In our visibility derivations we assumed 
the same effective wavelength $\lambda_{\rm eff}=2.2\,\mu$m
for all calibrators and science objects.
This effective wavelength results from the assumption
of a simple rectangular response function
for the $K$-band filter used
(central wavelength 2.2$\,\mu$m and bandwidth 0.4$\,\mu$m) 
and a constant spectrum of the calibrators/science objects
within the $K$~band.
This simplification introduces only a minor error in our calculations:
the effective wavelength $\lambda_{\rm eff}$
will be different from $2.2\,\mu$m since the filter
response function is not rectangular in shape (see, e.g., \citealt{psiphe}),
and the calibrator stars have diverse spectra.
Our calculations using the calibrator star with the earliest spectral type (Sirius) 
lead to an error of the calibrated object visibility of 0.23\% at 16\,m baseline 
(the largest baseline for the observations discussed in this paper).\\
The final errors (Fig.~\ref{vis}) of the calibrated visibilities were derived from 
simple Gaussian error propagation of the error of the fits to the measured transfer 
function values and the error of the uncalibrated visibility 
together with the additional effective wavelength error. 
\begin{table}[!ht]
\footnotesize
\begin{center}
\caption {List of calibrators used (see \citealt{CHARM}) and their UD angular diameters $d_{\rm{UD}}^{\rm a}$.}
\label{cal}
\begin{tabular}{l l l l}\hline
calibrators & $d_{\rm{UD}}^{\rm a}$ [mas] & nights & sp. type \\ \hline
30 Psc        & $7.20 \pm 0.70$  & 2001 Oct 22               & M3III   \\ 
$\tau$~Aqr    & $4.97 \pm 0.05$  & 2001 Oct 22,23            & K5III   \\ 
$\gamma$~Eri  & $8.51 \pm 0.09$  & 2001 Oct 22,23,           & M1IIIb  \\
              &                  & 2001 Nov 09,15,           &         \\
              &                  & 2001 Dec 06,              &         \\
              &                  & 2002 Dec 20               &         \\
$\alpha$~Cet  & $11.52 \pm 0.29$ & 2001 Oct 22,              & M1.5IIIa\\
              &                  & 2002 Dec 20               &         \\ 
$\chi$~Aqr    & $6.70 \pm 0.15$  & 2001 Oct 23               & M3III   \\
$\tau$~Pup    & $4.49 \pm 0.07$  & 2001 Oct 23,              & K1III   \\
              &                  & 2002 Jan 20,              &         \\
              &                  & 2002 Dec 20,              &         \\
              &                  & 2003 Jan 09               &         \\
$\pi$~Eri     & $4.80 \pm 0.50$  & 2001 Nov 09,15,17         & M1III   \\
Sirius        & $5.60 \pm 0.15$  & 2001 Nov 09,15,17,        & A1V     \\
              &                  & 2002 Jan 20,              &         \\
              &                  & 2002 Dec 20               &         \\
$\beta$~Cet   & $5.18 \pm 0.06$  & 2001 Oct 22,23,           & K0III   \\ 
              &                  & 2001 Nov 15,17,           &         \\
              &                  & 2001 Dec 06,              &         \\
              &                  & 2002 Dec 20               &         \\
$\delta$~CMa  & $3.29 \pm 0.46$  & 2001 Nov 15               & F8Iab   \\
HD~12274      & $5.30 \pm 0.50$  & 2002 Jan 20               & K5/M0III\\
$\theta$~CMa  & $4.13 \pm 0.40$  & 2002 Jan 20,              & K4III   \\
              &                  & 2002 Dec 20               &         \\
$\alpha$~Hya  & $9.44 \pm 0.90$  & 2002 Jan 20,              & K3II-III\\ 
              &                  & 2003 Jan 09               &         \\
$\pi$~Leo     & $4.78 \pm 0.26$  & 2002 Jan 20               & M2III   \\
$\eta$~Cet    & $3.35 \pm 0.04$  & 2002 Dec 20               & K1.5III \\
$\epsilon$~Lep& $5.90 \pm 0.06$  & 2001 Oct 22,23            & K4III   \\
              &                  & 2002 Dec 20,              &         \\
              &                  & 2003 Jan 09               &         \\
$\zeta$~Hya   & $3.10 \pm 0.20$  & 2003 Jan 09               & G9II-III\\
$\iota$~Hya   & $3.41 \pm 0.05$  & 2003 Jan 09               & K2.5III \\
$\mu$~Hya     & $4.69 \pm 0.50$  & 2003 Jan 09               & K4III   \\
$\beta$~Cnc   & $4.88 \pm 0.03$  & 2003 Jan 09               & K4III   \\ \hline
\end{tabular}
\end{center}
\end{table}
\begin{table*}[!htb]
\begin{center}
\caption { Properties of Mira model series from BSW, HSW, ISW, and TLSW (see text).
$L$ is the luminosity, $M$ the stellar mass, and $P$ the pulsation period in days. 
$R_{\rm p}$ is the Rosseland radius of the initial non-pulsating  ``parent'' model of the Mira variable, 
and $\alpha$ is the mixing-length parameter of convection (=the mixing-length
in units of the local pressure scale height, see \citealt{BOV}).
}
\label{models}
\begin{tabular}{c c c c c c c c}\hline
Series & Mode & $P$(d) & $M$/$M_{\odot}$ & $L$/$L_{\odot}$ & $R_{\rm p}$/$R_{\odot}$ & $T_{\rm{eff}}$/K & $\alpha$\\ \hline
D      & f    & 330    & 1.0             & 3470            & 236                     & 2900             & 1.76\\ 
E      & o    & 328    & 1.0             & 6310            & 366                     & 2700             & 1.26\\ 
P      & f    & 332    & 1.0             & 3470            & 241                     & 2860             & 2.06\\ 
M      & f    & 332    & 1.2             & 3470            & 260                     & 2750             & 1.73\\
O      & o    & 320    & 2.0             & 5830            & 503                     & 2250             & 0.93\\ \hline
\end{tabular}
\end{center}
\end{table*}
%
   \section{Comparison of the observations with Mira star models}\label{mods}
\subsection{Mira star models}\label{msm}
All Mira star models used in this paper are from BSW (D and E series), HSW (P, M, and O series), 
TLSW (P series), and from ISW (P series).
They were developed as possible representations of $o$ Cet and hence have periods $P$ close to 332 days.
These models differ in pulsation mode, stellar mass $M$, parent star radius $R_{\rm p}$ 
(i.e., radius of the initial non-pulsating model), and luminosity $L$.
The parent-star luminosities of the model series had been chosen based on considerations of the period-
luminosity-relation for the LMC and a theoretical metallicity correction (BSW, HSW).
Solar abundances were assumed for all models.
The five model series represent stars pulsating in the fundamental mode (\emph{f}; D, P, and M) or 
in the first overtone mode (\emph{o}; E and O). Table \ref{models} lists the properties of
the non-pulsating parent stars of these Mira models, 
and Table \ref{models2} shows the Rosseland radius and effective temperature associated with all P-series models.
All models are named after the cycle (see Fig. \ref{lightcurve} for the definition of cycle) 
and phase they represent. For example, P28 denotes the P model calculated
near visual phase 0.8 in the second cycle. The suffix ``n'', (e.g., P11n), denotes 
the new models from the recent fine phase grid of ISW.\\
\begin{table}[htb]
\begin{center}
\caption {
The variability phase $\Phi$, Rosseland radius $R_{\rm Ross}$ in units of the parent star radius $R_{\rm p}$,
$K$-band radius $R_K$ in units of the Rosseland radius,
radius $R_{1.04}$, which is defined by the position of the $\tau$=1 layer in the 1.04~$\mu$m 
continuum window in units of $R_{\rm p}$, the effective 
temperature $T_{\rm{eff}}(R)$, and the luminosity $L$ for all  P-series models from HSW and TLSW. 
Models denoted by ``n'' are from ISW with finer phase spacing (see text).
Phases in the 1+0.0 to 3+0.0 range were re-assigned as described in ISW.}
\label{models2}
\begin{tabular}{l @{\hspace{3mm}} l @{\hspace{3mm}} c @{\hspace{3mm}} c @{\hspace{3mm}} c @{\hspace{3mm}} c @{\hspace{3mm}} c}\hline
Model & cycle+$\Phi$  & $R_{\rm Ross}$/$R_{\rm p}$ & $R_K$/$R_{\rm Ross}$  & $R_{1.04}$/$R_{\rm p}$ & $T_{\rm{eff}}(R)$ & $L$\\ 
      &               &                            &                       &                        &   [K]             & [$L_{\odot}$]    \\\hline
P05    & 0+0.5                     &  1.20  &  0.65   & 0.90 & 2160 & 1650\\
P08    & 0+0.8                     &  0.74  &  0.97   & 0.74 & 3500 & 4260\\
P10    & 1+0.00                    &  1.03  &  0.94   & 1.04 & 3130 & 5300\\
P11n   & 1+0.10                    &  1.17  &  0.88   & 1.19 & 2990 & 5650\\
P12    & 1+0.23                    &  1.38  &  0.78   & 1.30 & 2610 & 4540\\
P13n   & 1+0.30                    &  1.53  &  0.69   & 1.26 & 2310 & 3450\\
P14n   & 1+0.40                    &  1.73  &  0.58   & 1.19 & 2080 & 2920\\
P15n   & 1+0.50                    &  1.88  &  0.49   & 0.84 & 1800 & 1600\\
P15    & 1+0.60                    &  1.49  &  0.47   & 0.85 & 1930 & 1910\\
P18    & 1+0.87                    &  0.77  &  0.94   & 0.77 & 3520 & 4770\\
P20    & 1+0.99                    &  1.04  &  0.92   & 1.04 & 3060 & 4960\\
P21n   & 2+0.11                    &  1.23  &  0.87   & 1.21 & 2790 & 4750\\
P22    & 2+0.18                    &  1.32  &  0.82   & 1.26 & 2640 & 4400\\
P23n   & 2+0.30                    &  1.36  &  0.78   & 1.24 & 2470 & 3570\\
P24n   & 2+0.40                    &  1.38  &  0.72   & 1.16 & 2210 & 2380\\
P25    & 2+0.53                    &  1.17  &  0.66   & 0.91 & 2200 & 1680\\
P28    & 2+0.83                    &  0.79  &  0.97   & 0.79 & 3550 & 5200\\
P30    & 2+0.98                    &  1.13  &  0.89   & 1.14 & 3060 & 5840\\
P35    & 3+0.5                     &  1.13  &  0.63   & 0.81 & 2270 & 1760\\
P38    & 3+0.8                     &  0.78  &  0.96   & 0.78 & 3570 & 5110\\
P40    & 4+0.0                     &  1.17  &  0.90   & 1.16 & 2870 & 4820\\ \hline
\end{tabular}
\end{center}
\end{table}
In this paper we use the conventional photospheric stellar Rosseland radius $R_{\rm Ross}$ 
given by the distance from the star's centre at which the
Rosseland optical depth equals unity.
This Rosseland radius is no observable
quantity (cf. the discussion in \citealt{BAS}; \citealt{SCH03}).
Usually, it is close to the near-infrared radius measured in near-continuum bandpasses
except at near-minimum phases with strong molecular bandpass contamination 
(HSW, \citealt{JS}, \citealt{SCH03}). 
Therefore, for better comparison Table~\ref{models2} also gives the $R_{1.04}$ radius 
that is defined by the position of the $\tau$=1 layer in the 1.04 $\mu$m 
continuum window.\\
In order to derive the model visibilites,
we averaged the $K$-band model CLV over all monochromatic CLVs 
between $2.0\,\mu$m and $2.4\,\mu$m.
An optimal comparison with VINCI visibilities, however, would require a slightly different
calculation of the polychromatic model visibility function:
for each monochromatic model CLV within the $K$~band, first  
the squared visibility function has to be calculated, and then
averaging over the entire wavelength range 
will deliver the correct squared visibility function. 
The square root of this average corresponds to
the visibility curve measured with VINCI (see, e.g., \citealt{psiphe}).
Comparison of the two methods for obtaining model visibility curves
shows only small discrepancies in the relevant spatial frequency range
and yields a systematic visibility difference of less than 0.5\%, which we included in 
the error budget of the calibrated visibility.\\
The Rosseland angular radii $R^{\rm a}_{\rm Ross}$ of $o$ Cet were determined in the following way: 
first, we derived the $K$-band angular radius $R^{\rm a}_K$, which is defined as the FWHM of the model CLV.
The $R^{\rm a}_K$ values (corresponding to different model-phase combinations) 
were determined by least-square fits between the measured Mira visibilities and
the visibilities of the corresponding model CLVs. 
Then, the Rosseland angular radii $R^{\rm a}_{\rm Ross}$ were derived from the obtained 
$K$-band angular radii and from the theoretical ratios $R_{\rm Ross}$/$R_K$ provided by the models.
For a more detailed description of the visibility fitting procedure we refer to HSW.

\subsection{Comparison of observed and model Rosseland linear radii}\label{radii}
\begin{figure*}[htb!]
  \centering
  \begin{tabular}{c @{\hspace{-3mm}} c @{\hspace{-3mm}} c}
    \includegraphics[angle=-90,width=6.2cm]{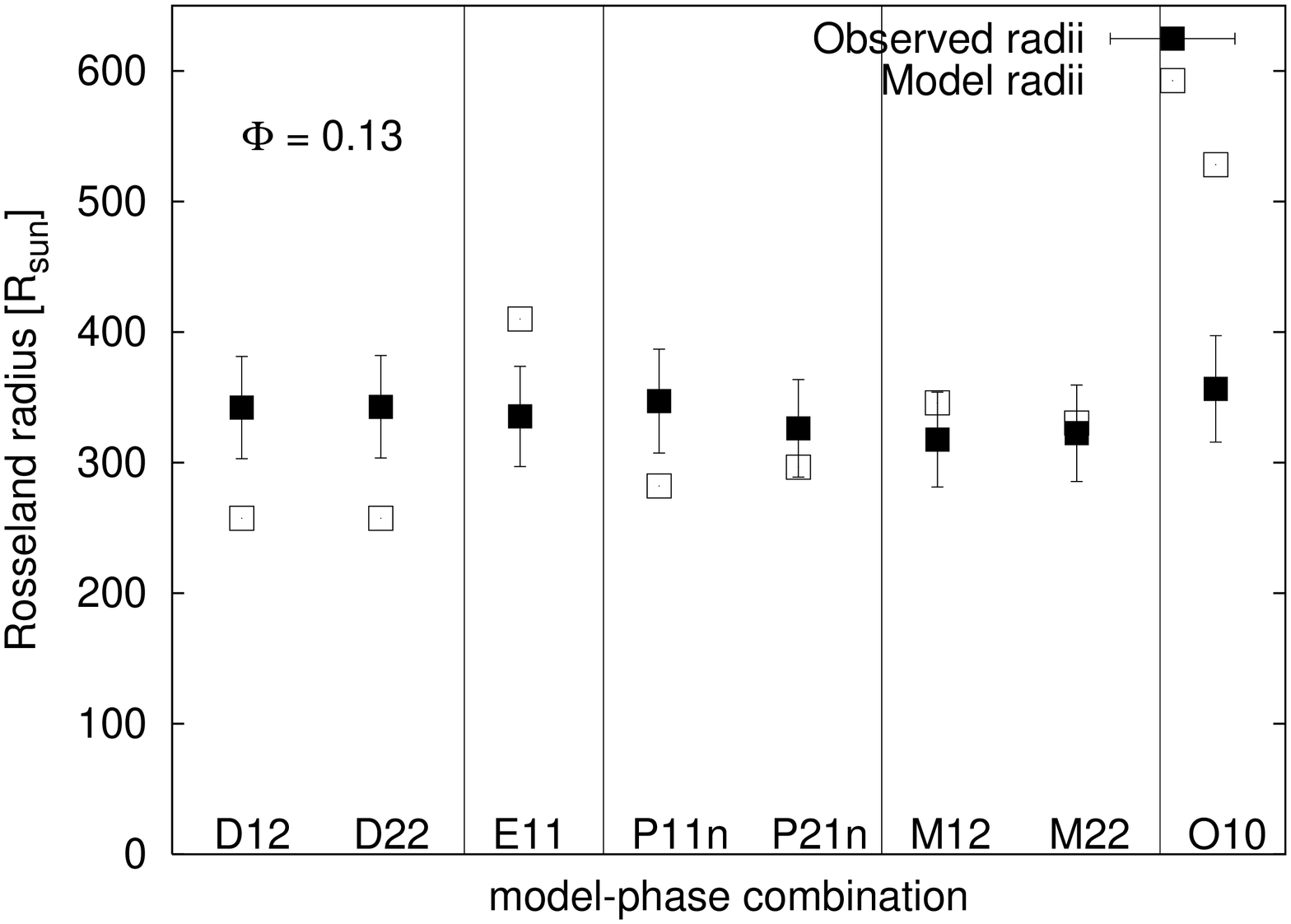}&
    \includegraphics[angle=-90,width=6.2cm]{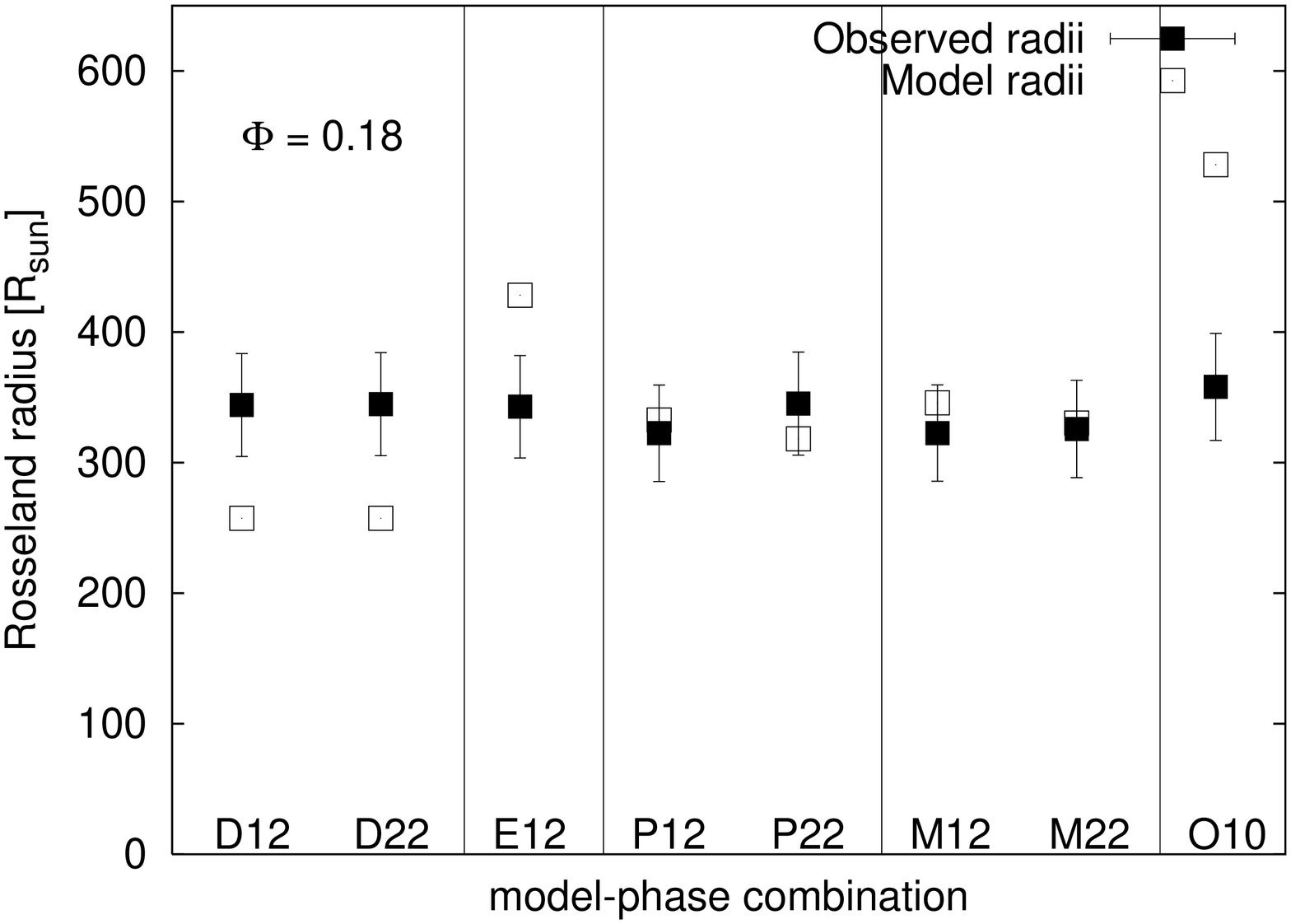}&
    \includegraphics[angle=-90,width=6.2cm]{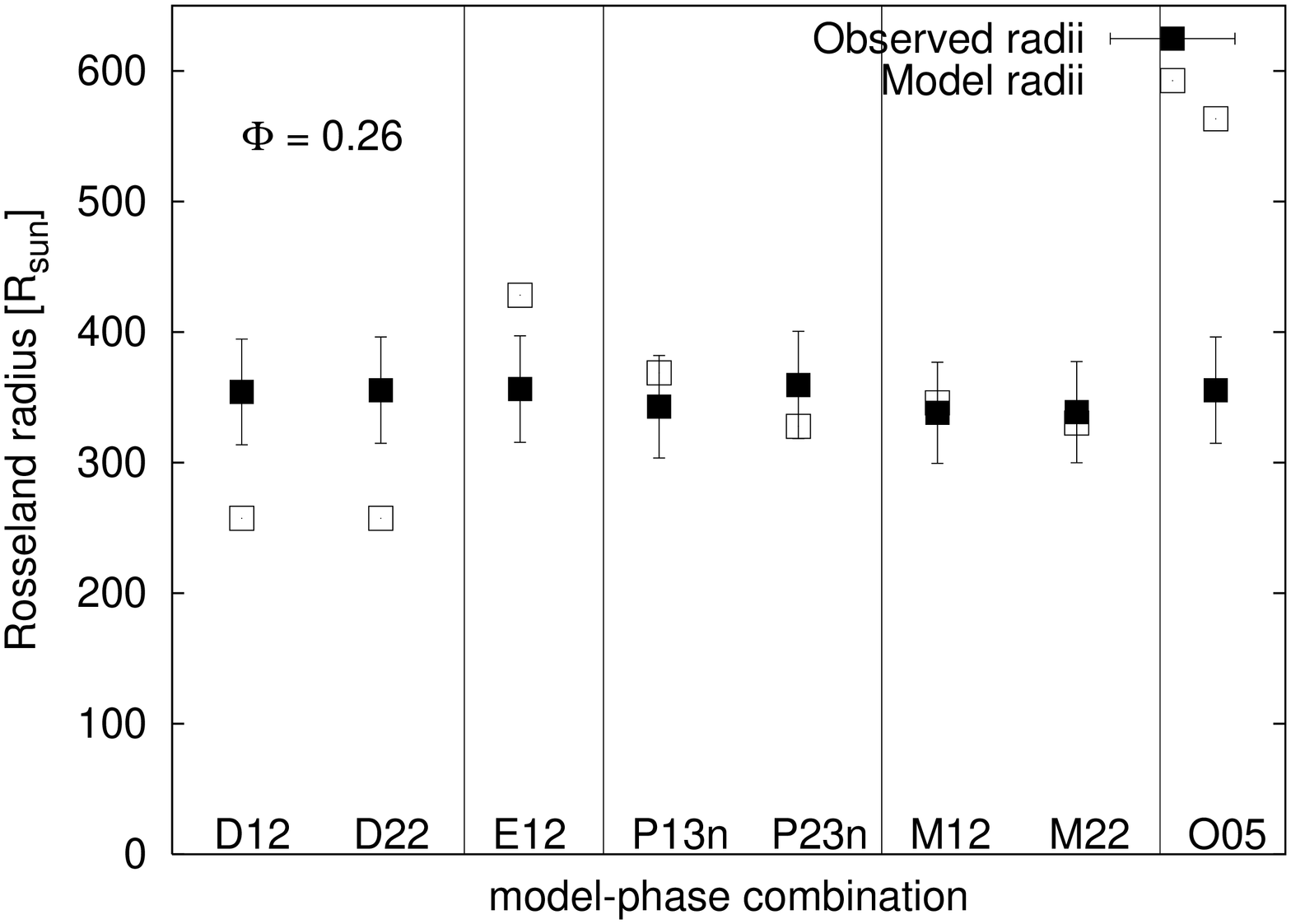}\\
    \includegraphics[angle=-90,width=6.2cm]{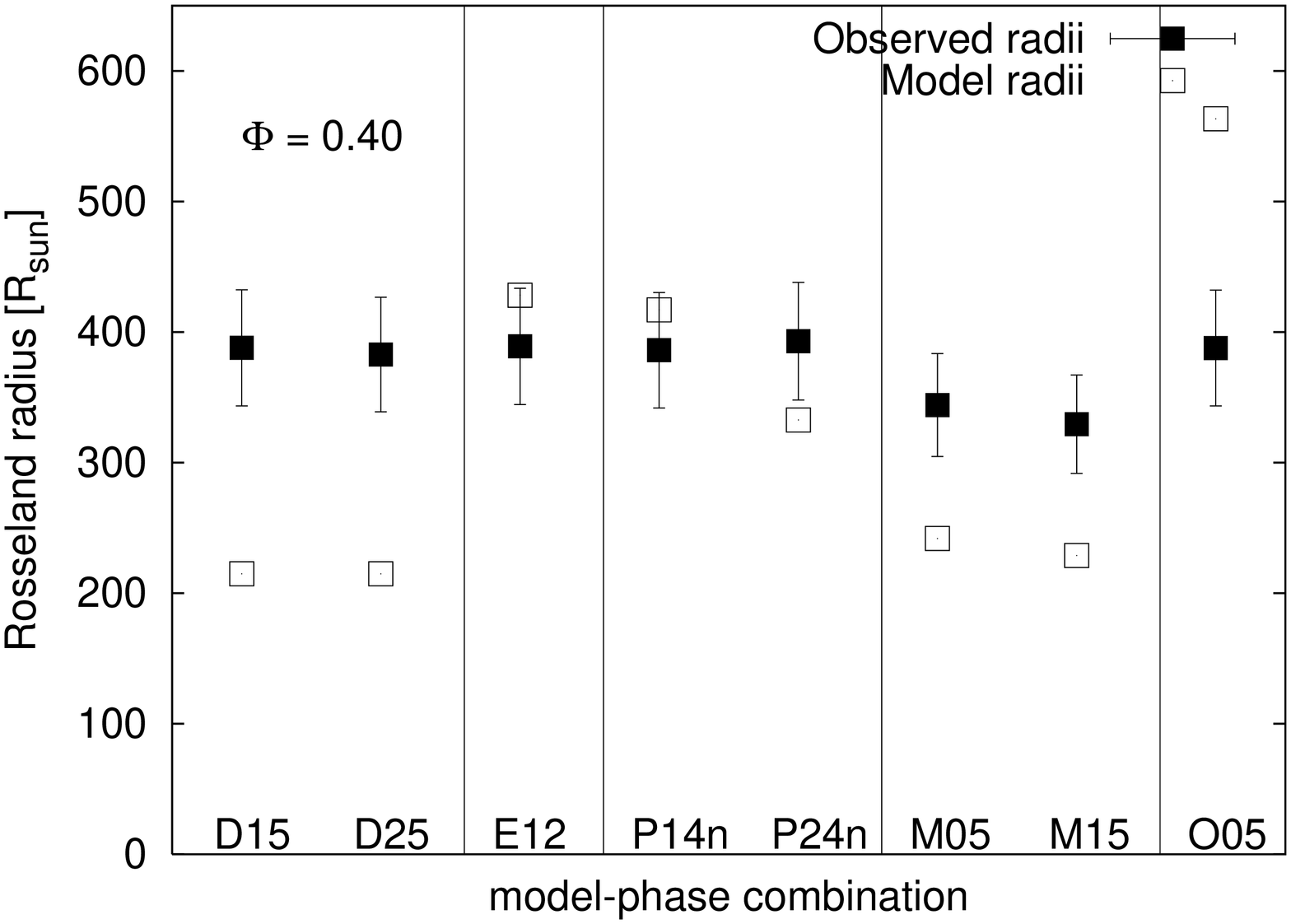}&
    \includegraphics[angle=-90,width=6.2cm]{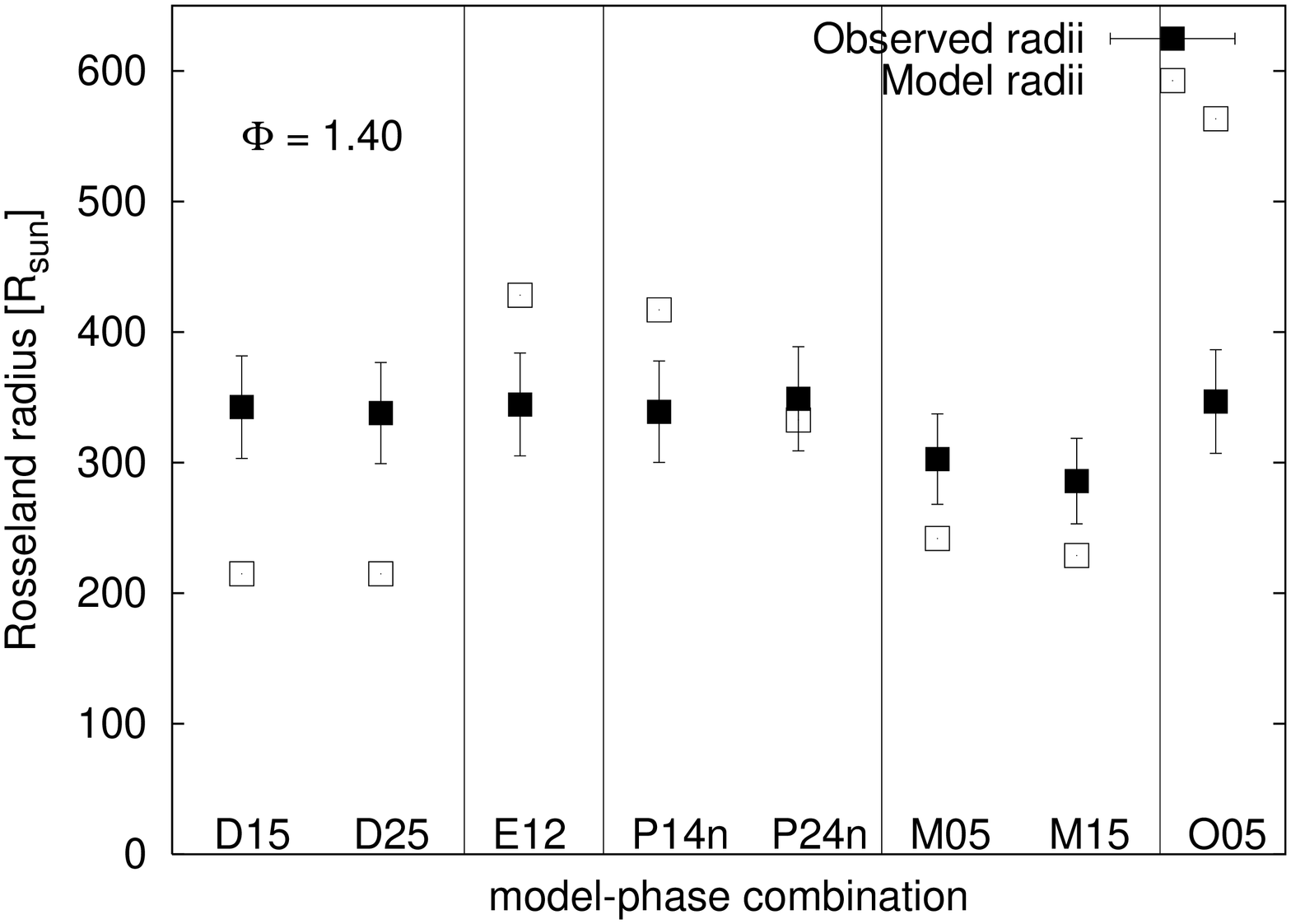}&
    \includegraphics[angle=-90,width=6.2cm]{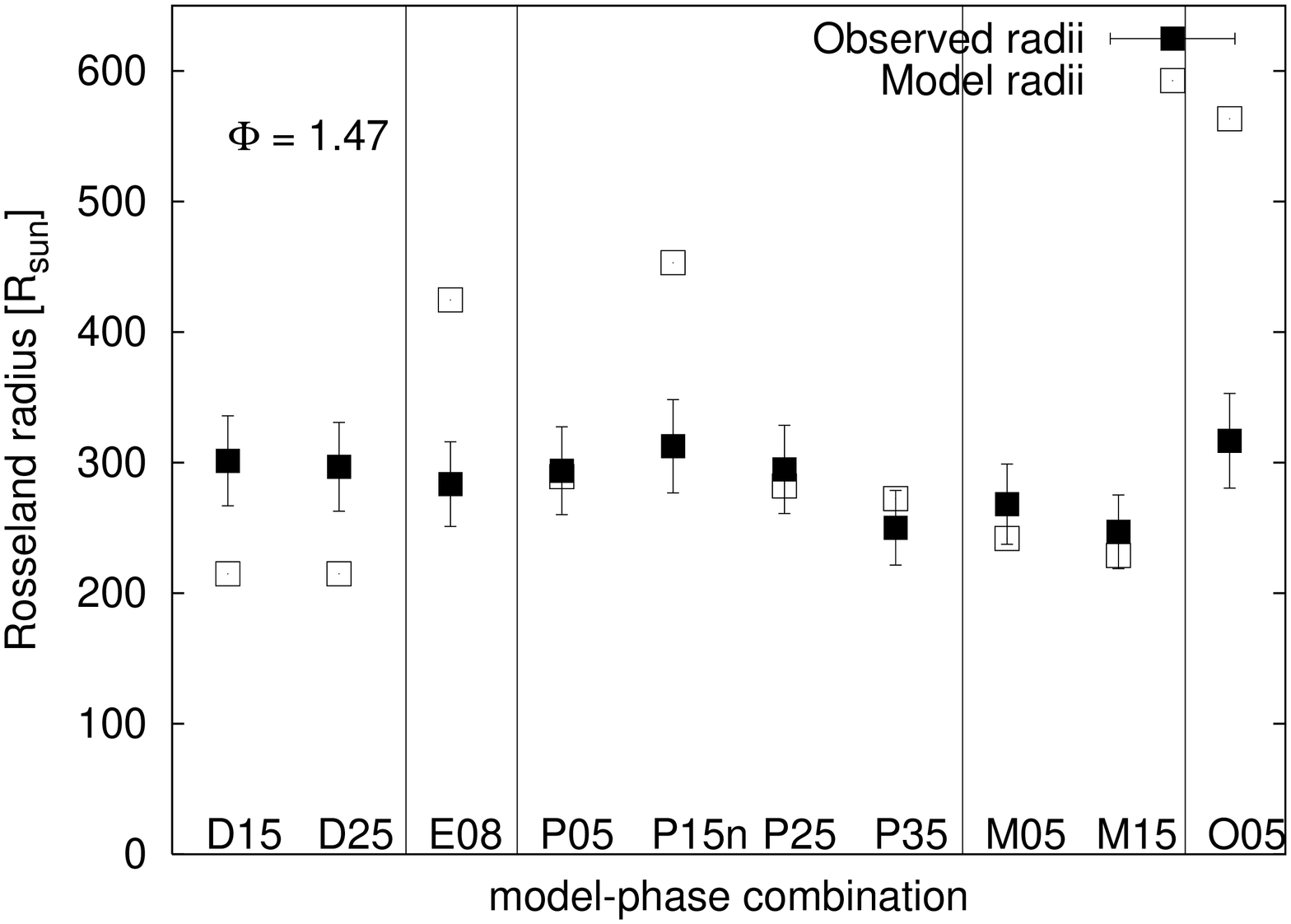}\\
  \end{tabular}
  \caption{Derived Rosseland linear radii for all epochs of observation and the Rosseland linear radii 
predicted by the nearest-phase models from all series. 
The Rosseland linear radii are given for the six measured phases $\Phi = 0.13$ (top left), 
$\Phi = 0.18$ (top right), $\Phi = 0.26$ (middle left), $\Phi = 0.40$ (middle right), 
$\Phi = 1.40$ (bottom left), and  $\Phi = 1.47$ (bottom right).
Open squares show the model values, while the filled squares with the error bars give the 
values derived from the observations.
The large error bars of $10-15\%$ are mainly due to the parallax error 
because the error contribution of the interferometric measurements is only approximately $1 \% $.}
  \label{figradii}
\end{figure*}

Linear Rosseland radii of $o$~Cet were obtained using the derived angular 
radii together with Mira's revised HIPPARCOS parallax of $9.34\pm 1.07$ mas (\citealt{KNAPP,ESA}).\\
Figure \ref{figradii} shows the comparison between the derived Rosseland linear radii and 
those of the corresponding models which provided the CLVs.
The $x$~axis represents  different combinations of model series, phase, and cycles.
\begin{figure}
  \centering
    \includegraphics[width=6cm,angle=-90]{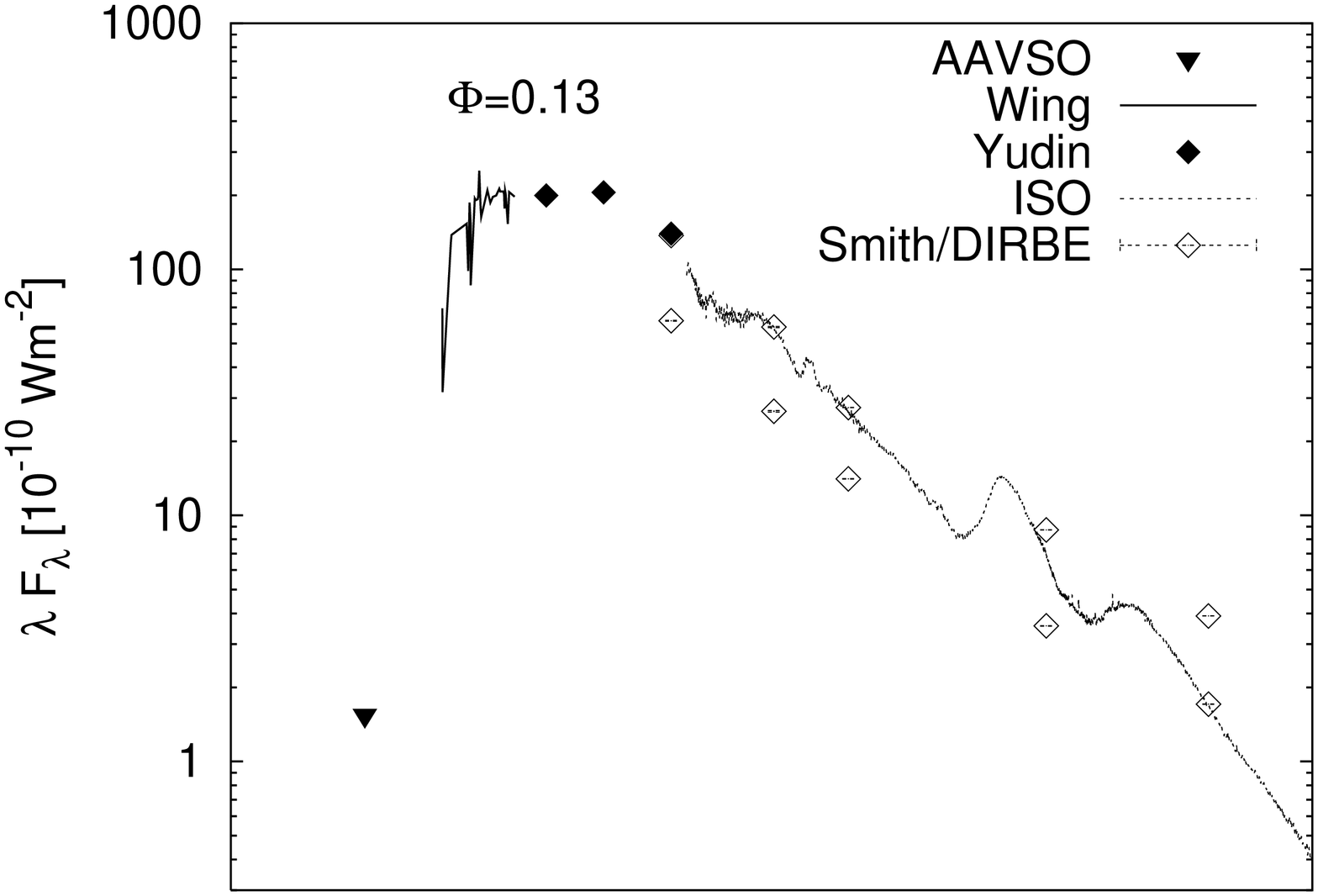}
\vspace*{-7.3mm}

    \includegraphics[width=6cm,angle=-90]{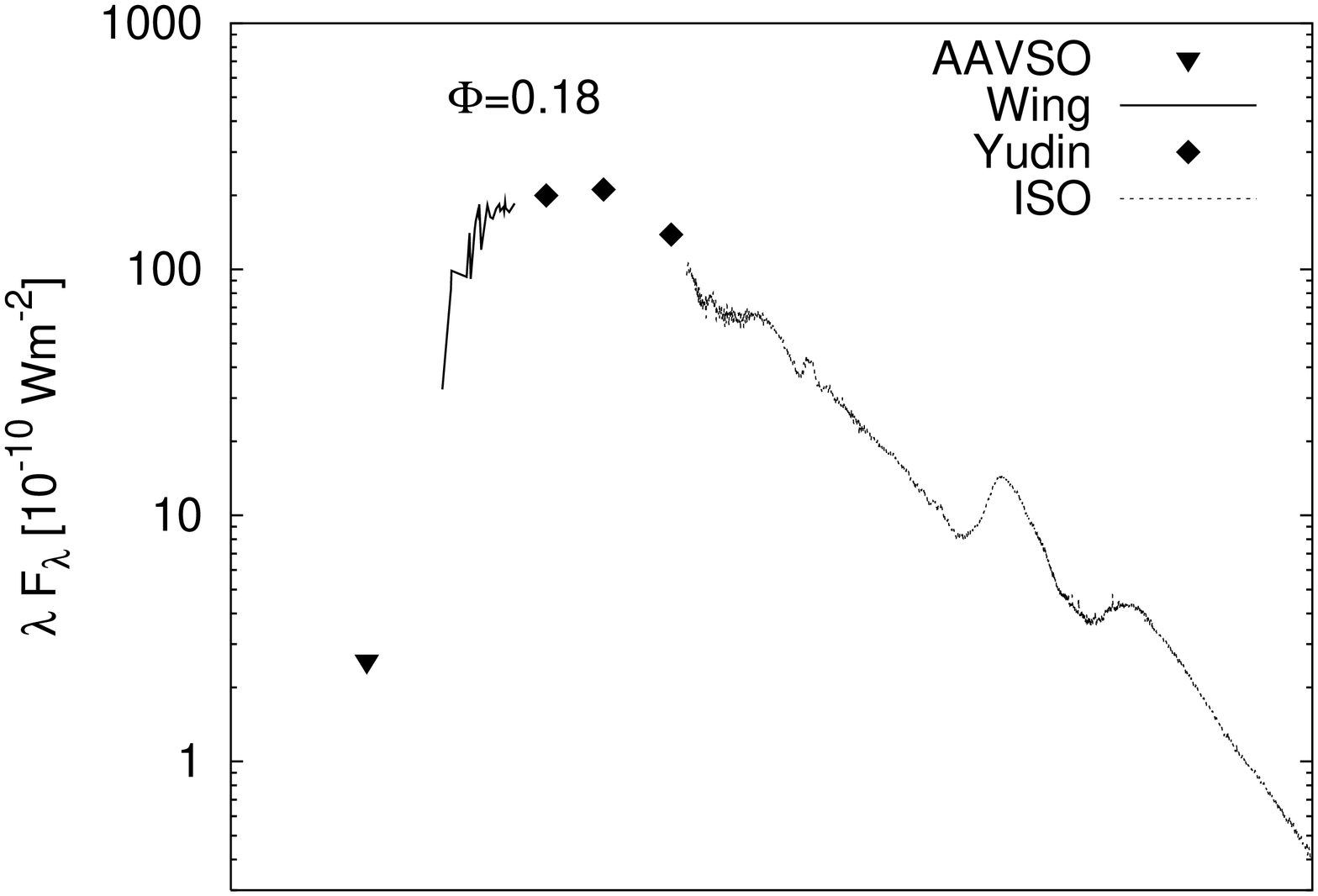}
\vspace*{-7.3mm}

    \includegraphics[width=6cm,angle=-90]{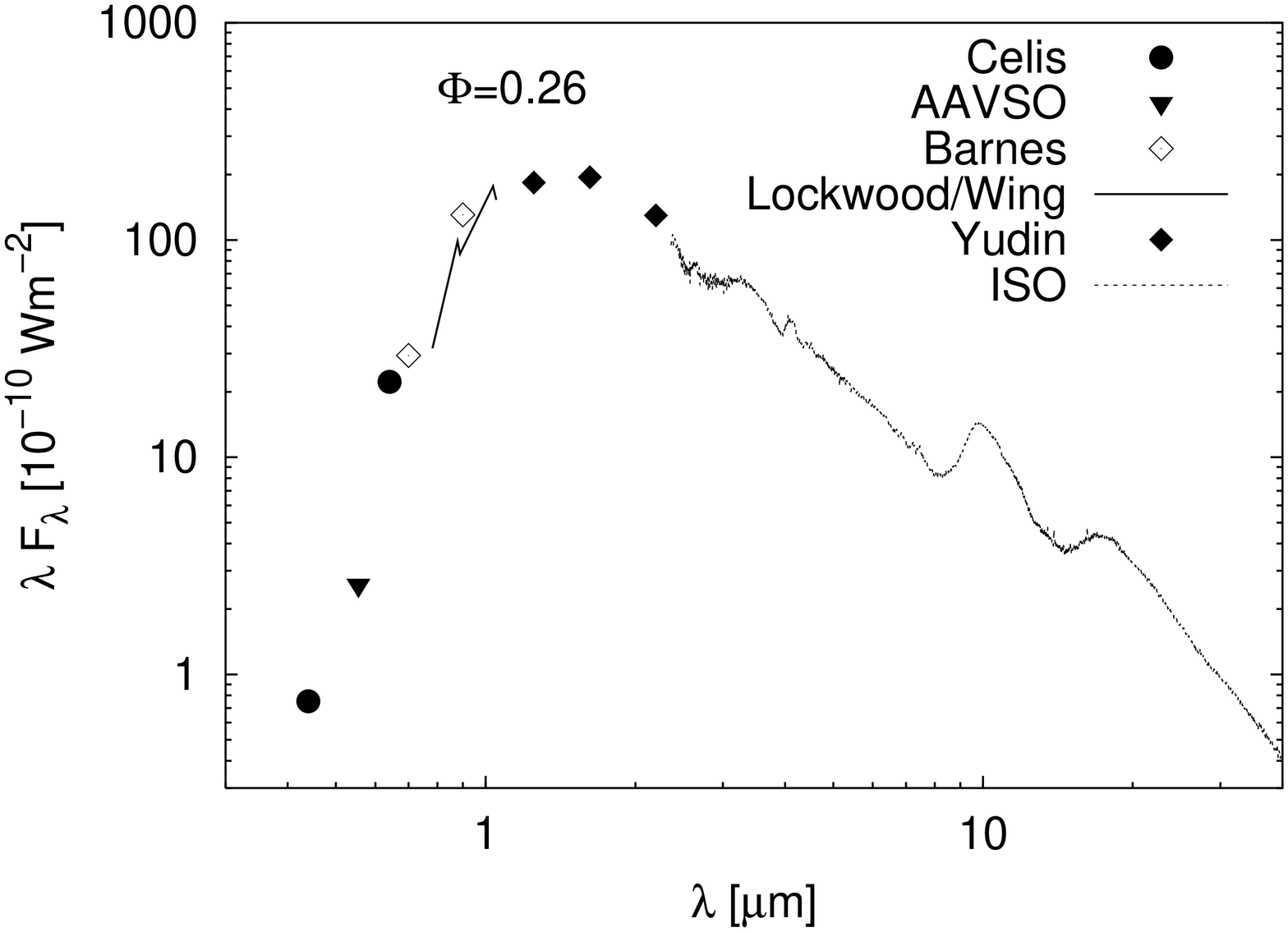}
\caption{Spectral energy distributions (SEDs) of $o$~Cet for phase $\Phi = 0.13$ (top), 
$\Phi = 0.18$ (middle), and $\Phi = 0.26$ (bottom).
The panels show visual flux densities from \cite{AAVSO} at the corresponding phase and cycle (triangles),
together with the ISO spectrum (longward of 2.36~$\mu$m, dotted line) taken at maximum light.
Also included are $JHK$ photometry measurements from \cite{BY} (filled diamonds) taken at 
the phase and cycle which correspond to the epoch of the VINCI observations.
The top and middle panels show spectrophotometry from \cite{RW} at 26 wavelengths 
between 0.78~$\mu$m and 1.10~$\mu$m (solid line) for the respective phases,
and the lower panel shows photometry data at 7 wavelengths: data at 0.44~$\mu$m and 
0.70~$\mu$m are from \cite{CEL82} (filled circles), 
and data between 0.78~$\mu$m to 1.05~$\mu$m are from \cite{LOCK71} (solid line).
Further measurements at 0.70~$\mu$m and 0.90~$\mu$m for $\Phi=0.23$ were taken from \cite{BAR} (open diamonds).
In the top panel we also plotted DIRBE flux densities taken at maximum and minimum light (open diamonds) 
\citep{SMITH02} in the 2.2, 3.5, 4.9, 12, and 25~$\mu$m photometric bandpasses.
}
\label{SED}
\end{figure}
\begin{figure}
  \centering
    \includegraphics[angle=-90,width=6.5cm]{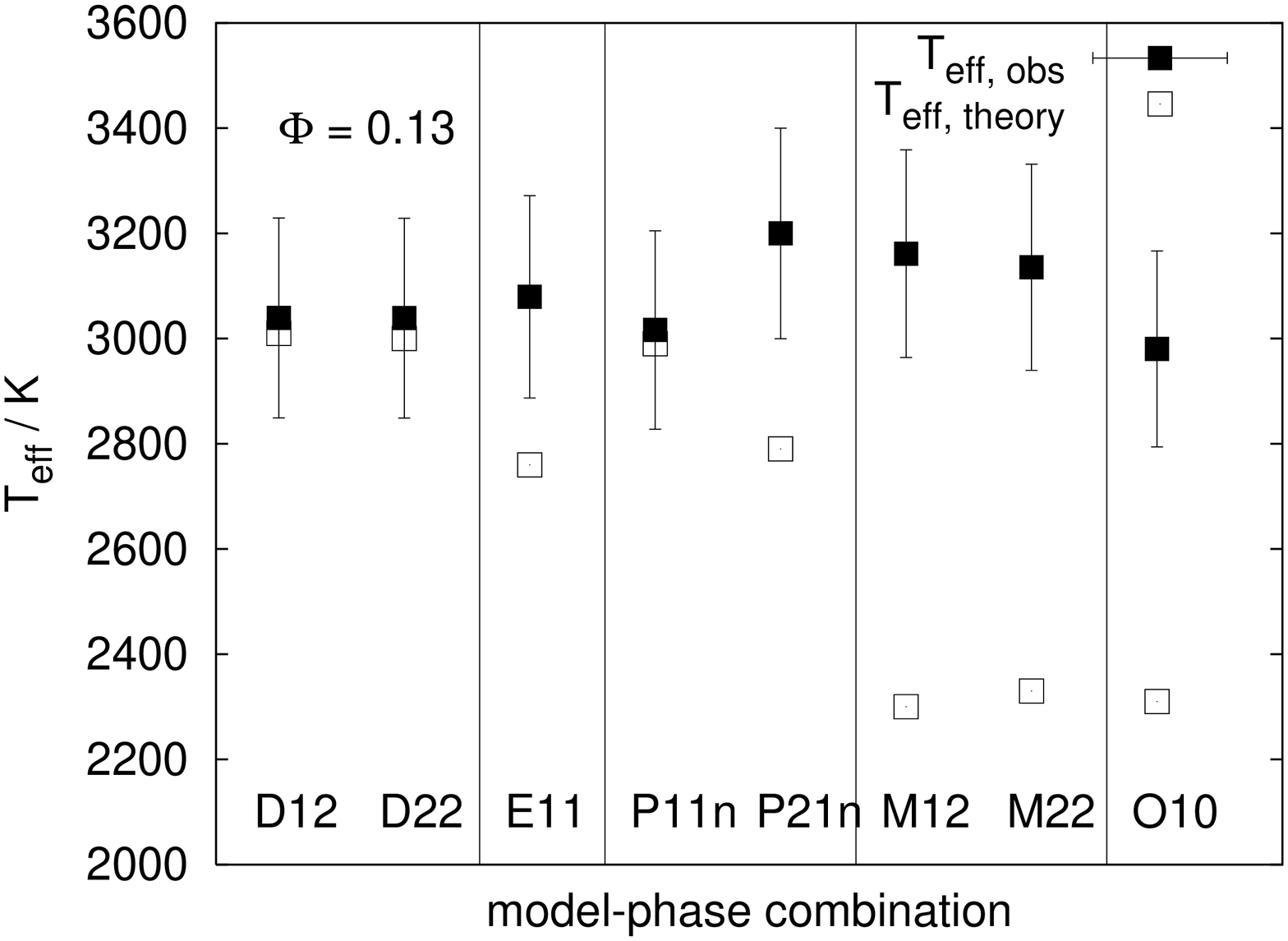}
    \includegraphics[angle=-90,width=6.5cm]{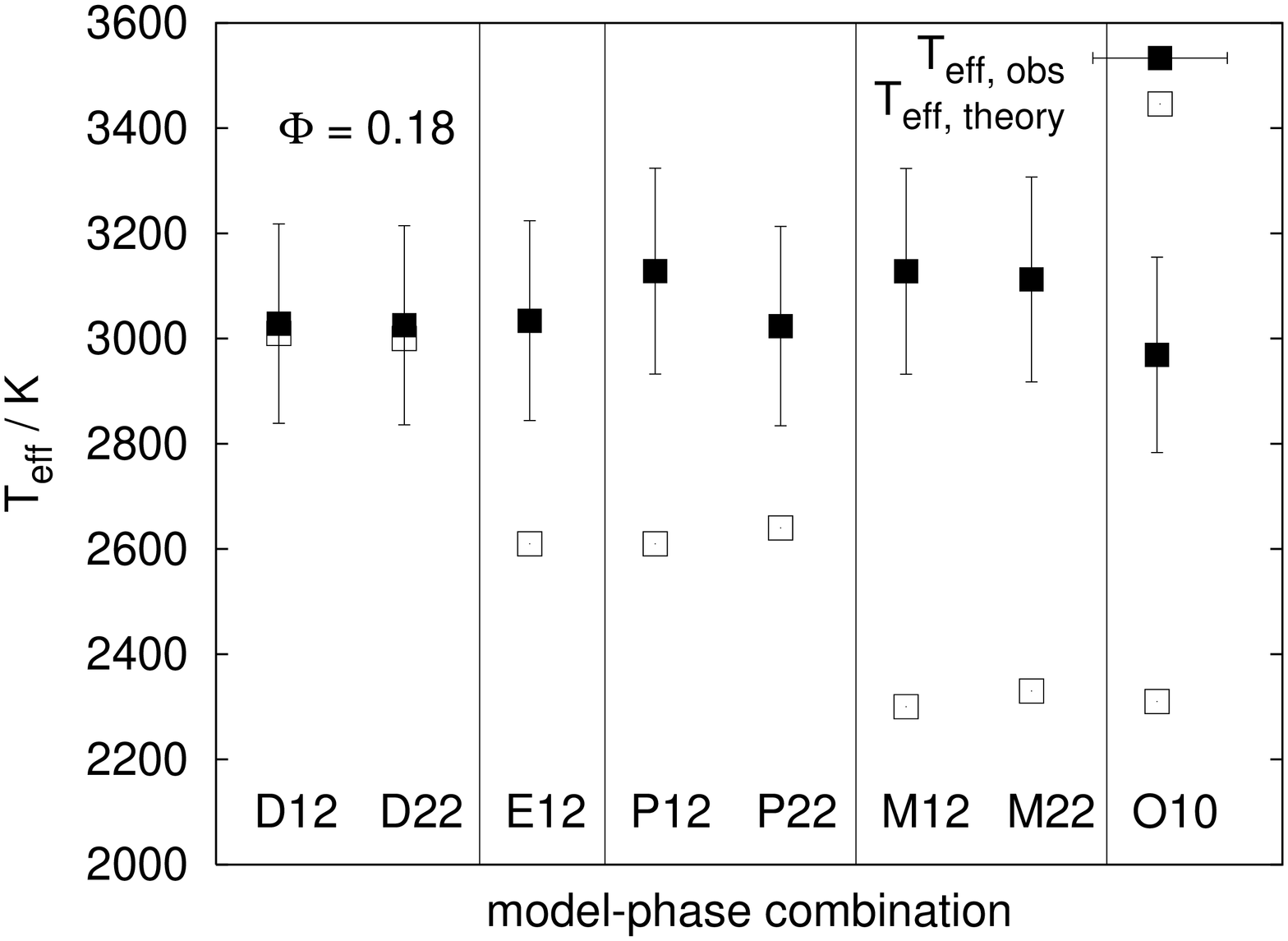}
    \includegraphics[angle=-90,width=6.5cm]{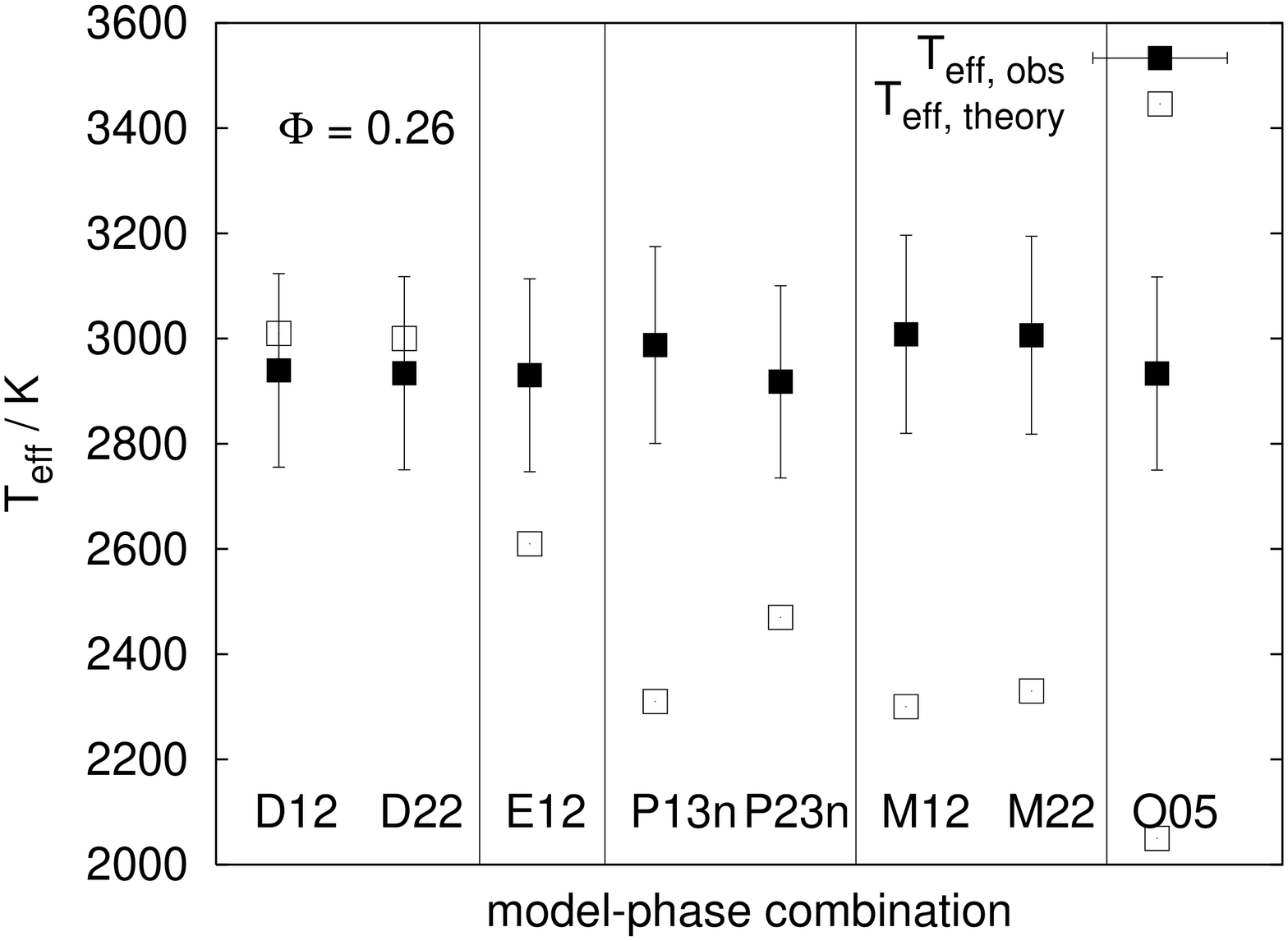}
\caption{Comparison of the effective temperatures derived from the observations using 
Eq. (\ref{teffeq}) at $\Phi=0.13,0.18$, and 0.26 (filled squares) 
and the model effective temperatures (open squares).
The measured effective temperatures were derived from the measured 
Rosseland angular radii (see Sect. \ref{radii}) and from our reconstructed SEDs.}
\label{tempmod}
\end{figure}
Clearly, the E and O first-overtone series and the D fundamental-mode series predict radii 
that are systematically either too small (D series) or too large (O and E series).
The fundamental-mode M series shows good agreement for the phases $\Phi =$ 0.13 - 0.26 and 1.47,
whereas the predicted radii are systematically too small at the phases 0.4 and 1.4.
The disagreement can be partially explained by the lack of M models for phases close to 0.4, 
since we use near-minimum (M05 and M15) models for the comparison. 
The absence of a fine phase grid also applies to the D and O model series, but agreement 
would probably not be improved with a finer phase grid, 
considering the large discrepancy between near-phase radii predictions and derived radii for 
$\Phi =$ 0.18,~0.26,~and~1.47.
The P-series models from HSW, TLSW, and ISW predict, in most cases (i.e., 10 out of 14 cycle-phase combinations), 
Rosseland linear radii which agree with the measured values within the error bars.
This good agreement of the P models is independent of the phase grid spacing.
The Rosseland linear radii predicted from the M models agree with the measured values 
(within the error bars) in 8 out of 12 cycle-phase combinations.
This radius comparison favors the fundamental-mode pulsation P and M models as possible representations of Mira,
whereas all first overtone pulsation model series can be ruled out as models that reliably predict 
the Rosseland radius of $o$~Cet.\\
\subsection{Determination of $T_{\rm eff}$}\label{teff}
The effective temperature $T_{\rm eff}$ of a star can be derived from its 
Rosseland angular diameter and its bolometric flux (cf. \citealt{ST};
\citealt{BAS}; \citealt{SCH03} for the issue of $T_{\rm eff}$ definition).
Using convenient units, the relation between $T_{\rm eff}$, $F_{\rm{bol}}$, and 
Rosseland angular diameter $d_{\rm Ross}^{\rm{a}}$ can be written as follows:
\begin{equation}
\label{teffeq}
T_{\rm eff} = 2341\left(\frac{F_{\rm bol}}{10^{-11}\rm{W~m}^{-2}}\right)^{\frac{1}{4}}
\left(\frac{d_{\rm Ross}^{\rm{a}}}{\rm mas}\right)^{-\frac{1}{2}} \rm{K}.
\end{equation}
In order to estimate the bolometric flux, we used photometric measurements of $o$ Cet in the visual
(\citealt{AAVSO}) and, when available, in the NIR in combination with the ISO 
spectrum ranging from 2.36 $\mu$m to 45.38 $\mu$m.
In the region between 0.7812 $\mu$m and 1.0834 $\mu$m spectrophotometric data were taken 
from measurements of \cite{RW}.
The 1.25 to 2.2 $\mu$m data were taken from \textit{JHK} photometry by \cite{BY}.
Since the ISO spectrum was taken at maximum light ($\Phi = 0$), 
we used light curves \citep{SMITH02} obtained with the Diffuse Infrared Background Experiment 
(DIRBE, see \citealt{SMITH03})  to estimate the variability of $o$ Cet in the ISO spectral range.
DIRBE light curves are available in seven infrared bands between 1.25 $\mu$m and 60 $\mu$m, 
giving us important complementary photometric data. Additional data used and the overall shape of 
the spectral energy distributions (SEDs) can be seen in Fig. \ref{SED}.\\
The light curves in the ISO range show that flux variations 
in the mid- and far-infrared between maximum and minimum light is small but not negligible. 
Since most of the bolometric flux comes from the NIR, 
the error of $F_{\rm bol}$ induced by flux density changes in the ISO range makes a contribution 
of only $\sim 5-10\%$ to the total error. Because of the small interstellar extinction in the $V$ and 
$K$ bands ($A_V = 0.09$, \citealt{ROBF}, and $A_K = 0.05$, \citealt{KNAPP}) 
and the proximity of Mira, no correction for interstellar extinction was applied.
Integration of the SED was made by simple trapezoid integration and yields 
$F_{\rm bol} = 2.63\cdot 10^{-8}$ Wm$^{-2}$ at $\Phi=0.13$, 
$F_{\rm bol} = 2.41\cdot 10^{-8}$ Wm$^{-2}$ at $\Phi=0.18$, 
and $F_{\rm bol} = 2.36\cdot 10^{-8}$ Wm$^{-2}$ at $\Phi=0.26$.
The total error of the bolometric flux was estimated to be 25\%. 
The error sources are mainly the cycle and phase uncertainties of the spectrophotometric 
and photometric observations. Since spectrophotometric and photometric data were not sufficiently 
available for reconstructing SEDs at phases later than 0.3,
we derived $T_{\rm eff}$ from the Rosseland angular diameter 
$d_{\rm Ross}^{\rm{a}}$ and the SED according to Eq.~(\ref{teffeq}) only for the phases
$\Phi = 0.13,~0.18$, and $0.26$. The error in $T_{\rm eff}$ is given by
\begin{equation}
\frac{\Delta T_{\rm eff}}{T_{\rm eff}}=\frac{1}{2}\left[\left(\frac{\Delta d_{\rm Ross}^{\rm{a}}}
{d_{\rm Ross}^{\rm{a}}}\right)^2+\frac{1}{4}\left(\frac{\Delta F_{\rm bol}}{F_{\rm bol}}\right)^2\right]^{\frac{1}{2}}
\end{equation}
\label{dteff}
and is $6\%$ when errors of approximately $25\%$ and $1 \% $ are assumed for 
$F_{\rm bol}$ and $d_{\rm Ross}^{\rm{a}}$, respectively.\\
Figure \ref{tempmod} presents $T_{\rm eff}$ values derived at phases close to the VINCI observations from the
bolometric flux and the Rosseland angular diameters.
The $T_{\rm eff}$ values of the D model series are very close (within the error bars) 
to those derived from the observations at all phases.
Although the P11n model shows agreement between observation and theory for phase 0.13, 
all other cycle-phase combination of the P-model series as well as all other
model series yield large differences between theoretical and measured
effective temperature values.\\
We also determined the luminosity range of $o$~Cet for the observed phases
from its bolometric flux and HIPPARCOS parallax (9.34 $\pm$ 1.07 mas, \citealt{KNAPP}).
We obtained $9360\pm 3140~L_{\odot}$ for $\Phi=0.13$ and $8400\pm 2820~L_{\odot}$ for $\Phi=0.26$.
These luminosities are higher than those predicted by all models for the respective phases 
except for the E model series, for which a higher luminosity of the parent star was adopted 
(see Table \ref{models}). 

\subsection{CLVs and visibility shapes}\label{visshape}
As discussed in Sect.~\ref{radii}, there is good agreement between the derived 
Rosseland linear radii and the Rosseland linear radii predicted by model series P and M~series models, 
while there is little or no agreement for the other models. 
On the other hand, the effective temperatures derived from observations are close to those 
predicted by the D-model series. 
In this section we investigate whether one of the model series can be favored by comparing the 
observed and theoretical visibility shapes.\\
The measured visibilities at six different phases within the first and second cycle of the 
observations are shown in Fig.~\ref{vis}, 
together with fitted model visibility curves for the P, M, and D models.
The P-series fundamental pulsator mode model P21n shows much better agreement with the 
observation at $\Phi=0.13$ than the D and M models 
or simple UD model CLVs. This can be clearly seen from the inset in Fig.~\ref{vis},
and a comparison of the reduced $\chi^2$ values for the different model fits. 
For model P21n we derived $\chi^2_{\rm P21n}=6.7$, while we obtained 
$\chi^2_{\rm D22}=41.5$ and $\chi^2_{\rm M22}=28.8$ for the models D22 and M22, respectively.
For the simple uniform-disk model, which is also shown in Fig.~\ref{vis}, we found  $\chi^2_{\rm UD}=56.2$.
For all other phases shown in Fig.~\ref{vis}, 
the small number of visibility points and the insufficient 
coverage of the spatial frequency domain
make it impossible to draw conclusions about which model series should be favored.
It should be noted that the agreement of the observed CLV shape with the P-series model 
CLV shape is also better than for the E and O first-overtone model series
(reduced $\chi^2$ values of 22.9 and 45.4, respectively), 
which were already ruled out from  radius and temperature comparisons.\\ 
The visibility can also be affected by a circumstellar dust shell 
which is not included in the model calculations.
A dust shell could cause a narrow visibility peak at low spatial frequencies.
Therefore, we measured the visibilities of $o$~Cet
using our speckle camera at the SAO 6~m telescope.
These additional visibilities cover the low spatial frequency domain, showing that
the use of dust-free CLV models is justified (see Fig.~\ref{vis}, middle left).
\\
Although the measured visibility shape for $\Phi=0.13$ is in good agreement with 
the predictions made by the P21n model, 
we nevertheless wish to point out that the observed visibility shape can also be affected by 
deviations from circular symmetry of the intensity distribution. 
Therefore, we studied possible effects of such asymmetries on the 
visibility shapes by assuming elliptical uniform-disk intensity distributions with ratios of 
minor to major axis between 0.7 and 1.0. 
It turned out that 
the observed visibilities cannot be reproduced by such elliptical
uniform disk intensity distributions.
However, we note that this study cannot 
exclude the presence of other types of asymmetries in Mira,
such as surface inhomogeneities produced by giant convective cells as suggested by \cite{SCHW}.

\subsection{Diameter-phase relationship}\label{diam}
\begin{figure}
\centering
\includegraphics[angle=-90,width=6.5cm]{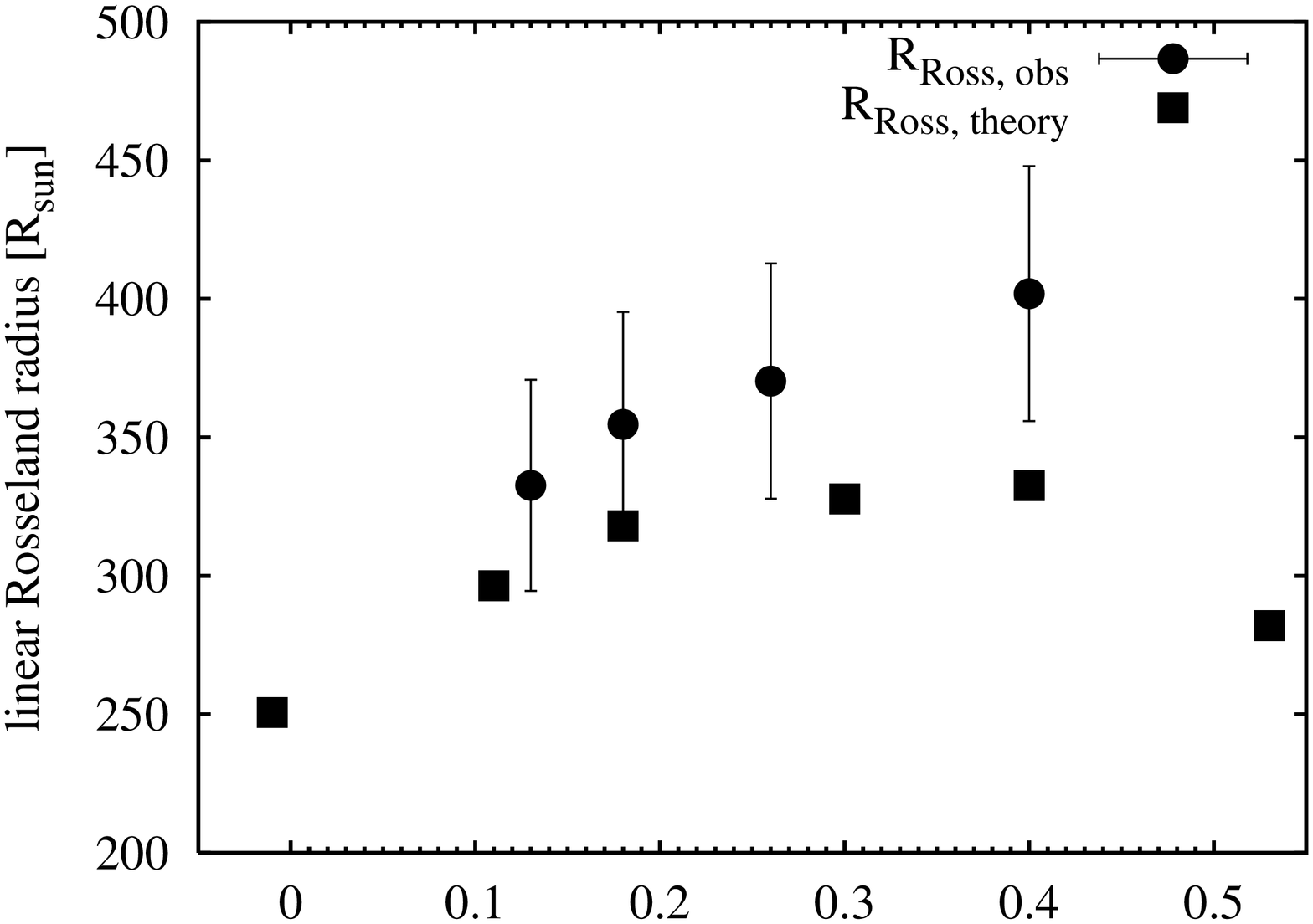}
\includegraphics[angle=-90,width=6.5cm]{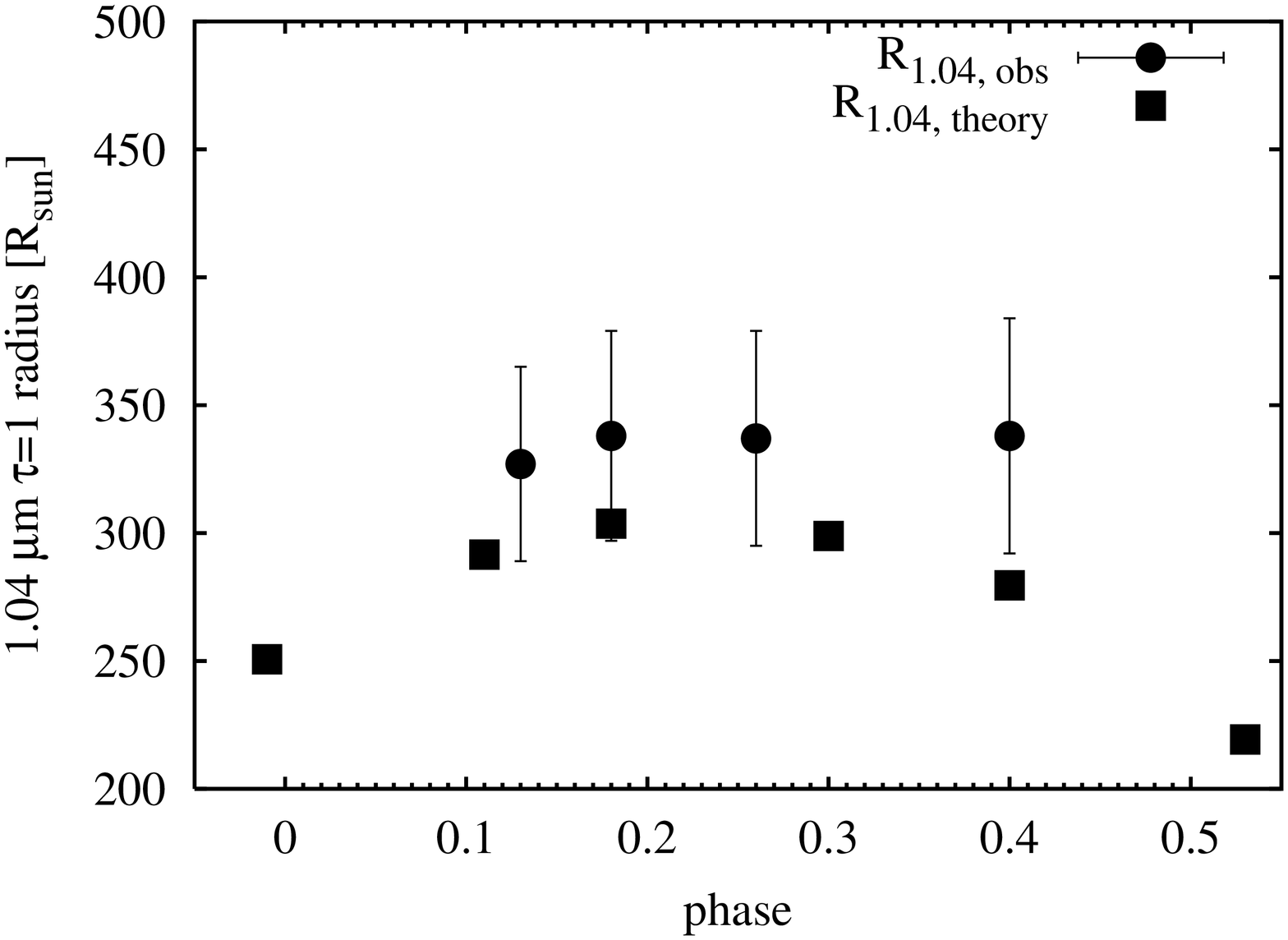}
\caption{{\bf Top}: Linear Rosseland radii of $o$~Cet derived from P21n, P22, P23n, and P24n 
series models (filled circles) at four different phases
($\Phi = 0.13$: $R_{\rm{Ross}} = 332 \pm 38$~$R_{\odot}$; $\Phi = 0.18$: $R_{\rm{Ross}} = 354 \pm 41$~$R_{\odot}$;
$\Phi = 0.26$: $R_{\rm{Ross}} = 370 \pm 42$~$R_{\odot}$; $\Phi = 0.40$: $R_{\rm{Ross}} = 402 \pm 46$~$R_{\odot}$)
and theoretical P2$x$ models (filled squares).
The diagram clearly shows a monotonic increase of radius with decreasing brightness, 
in line with the predictions from TLSW.
{\bf Bottom}: Derived (filled circles) and theoretical (filled squares) $R_{1.04}$ radii of $o$~Cet.}
\label{RossRad}
\end{figure}
\begin{figure*}
  \centering
  \begin{tabular}{c  c}
    \includegraphics[angle=-90,width=9.1cm]{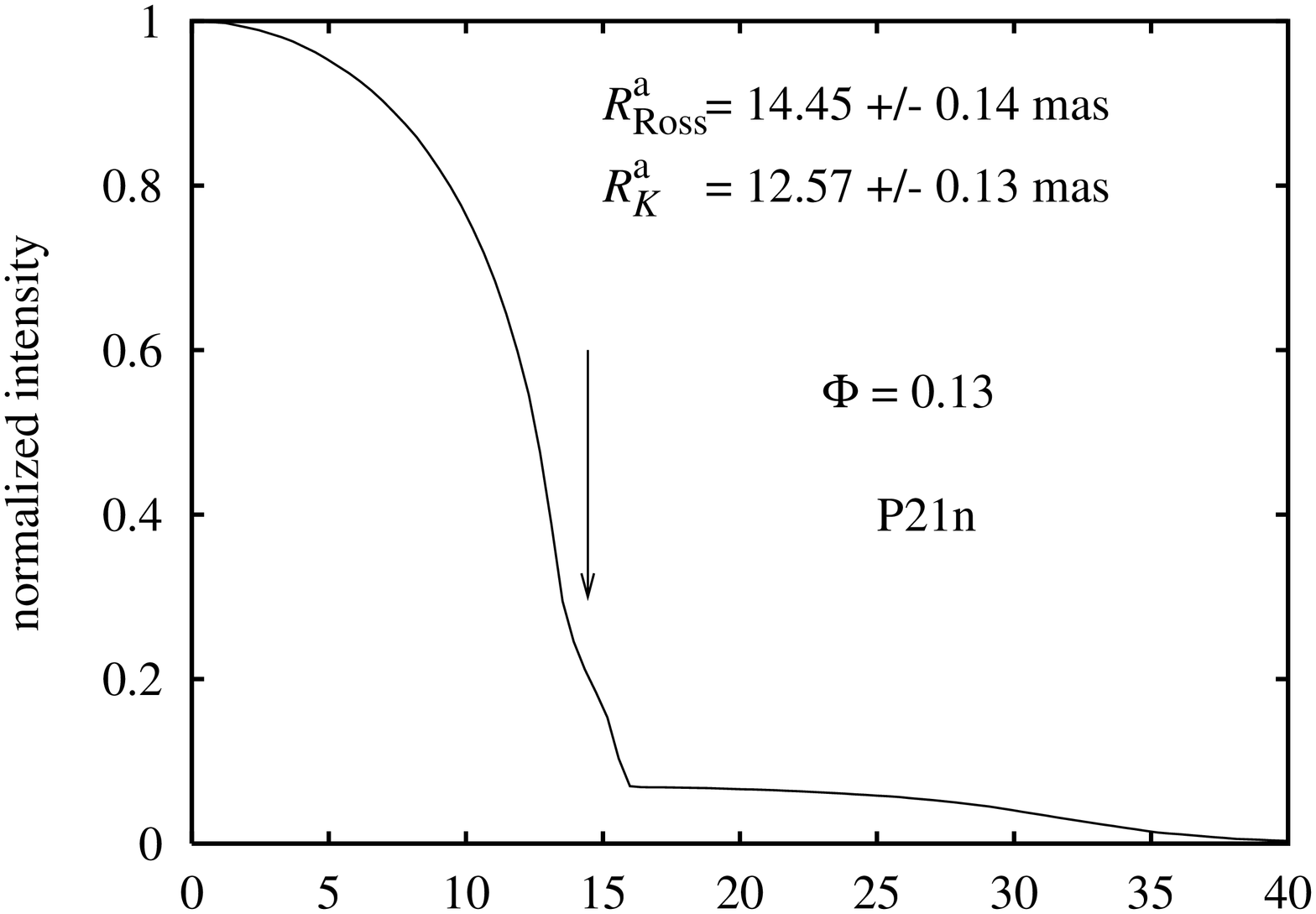}&
 \hspace{-1cm}   \includegraphics[angle=-90,width=9.1cm]{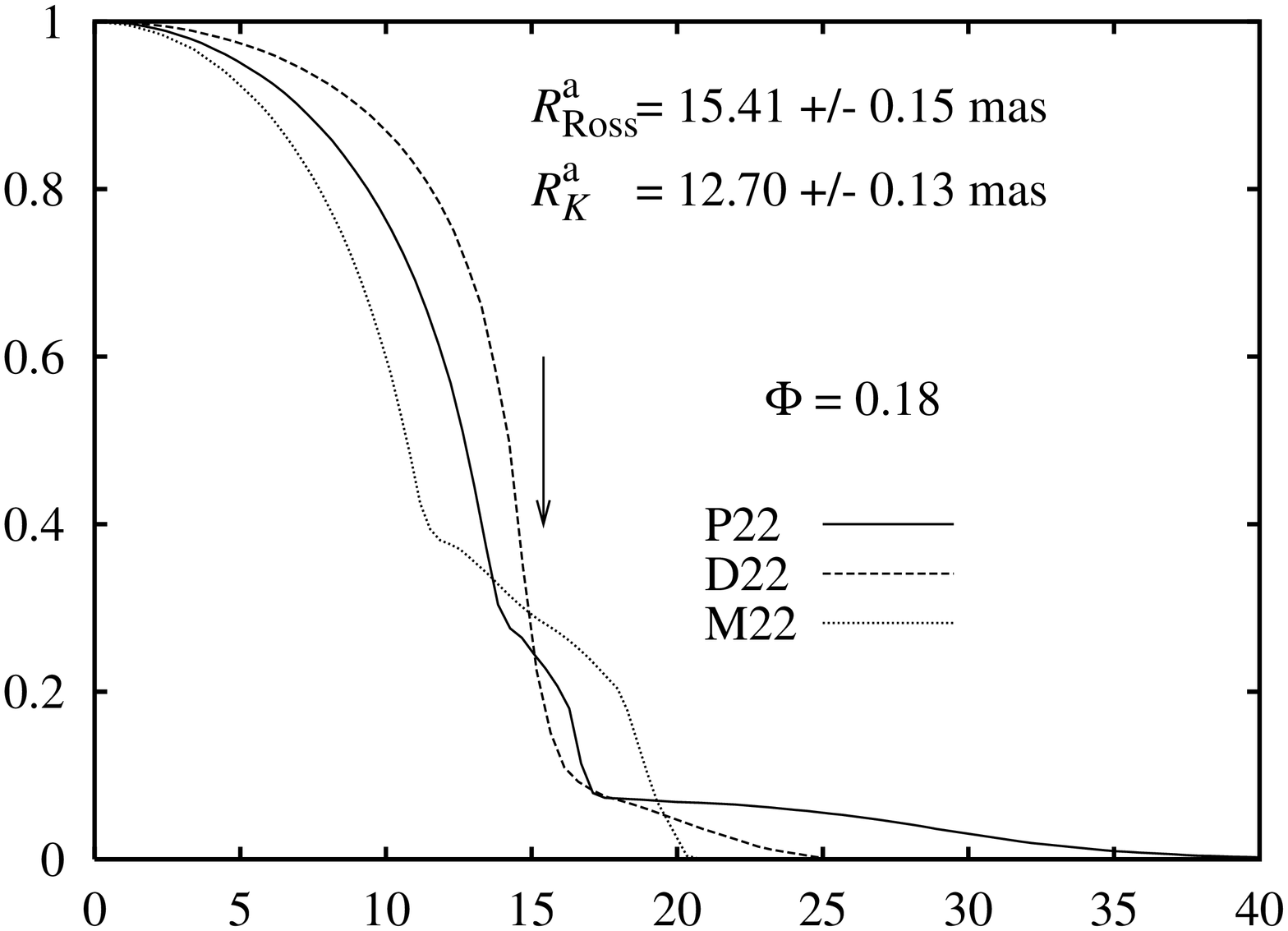}\\
    \includegraphics[angle=-90,width=9.1cm]{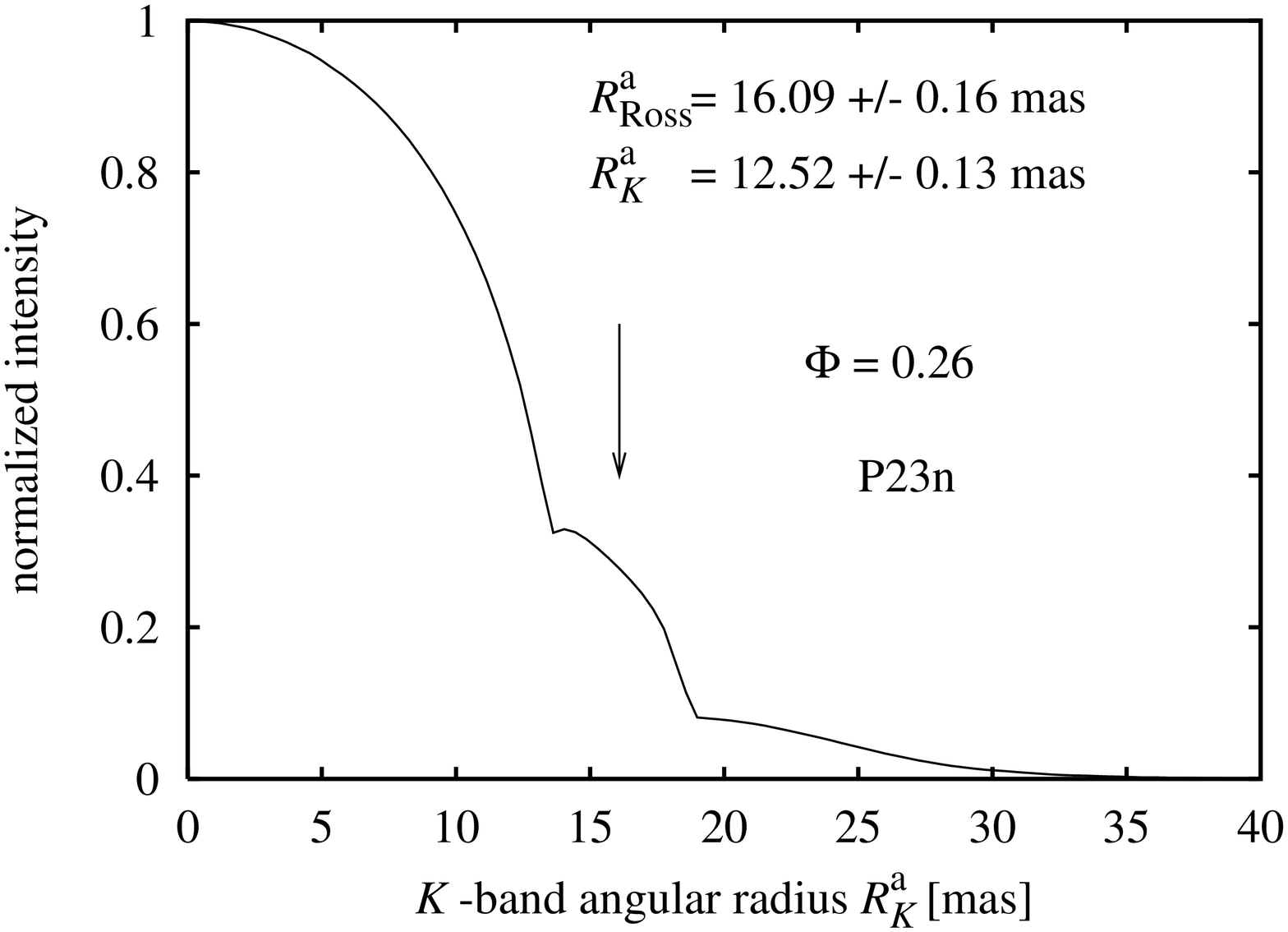}&
   \hspace{-1cm} \includegraphics[angle=-90,width=9.1cm]{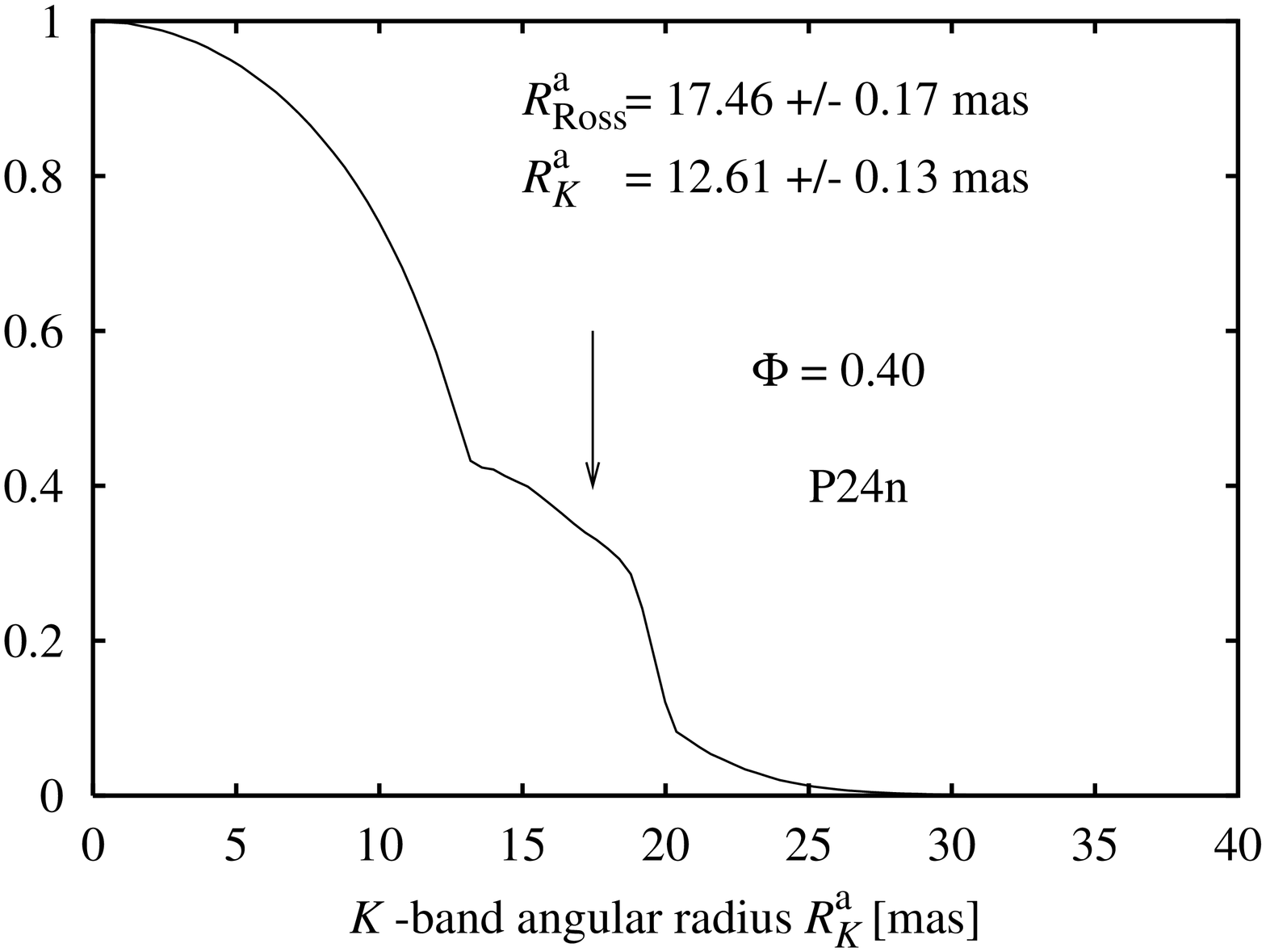}\\
  \end{tabular}
  \caption{
Theoretical CLV shapes of the ISW model P21n plotted for the $K$-band angular radius $R_K^a$ corresponding
to $o$~Cet at $\Phi=0.13$ ({\bf top left}), of the TLSW model P22 at $\Phi=0.18$ ({\bf top right}), 
of the ISW model P23n at $\Phi=0.26$ ({\bf bottom left)}, and of the ISW model
P24n at $\Phi=0.4$ ({\bf bottom right}) as a function of angular radius. 
The intensity is normalized to 1 at the disk center. 
For comparison, the intensity profiles of BSW model D22 ({\bf top right}, dashed line) 
and the HSW model M22 ({\bf top right}, dotted line) at $\Phi=0.18$ are also shown. 
The extended wing of the CLV for P models near maximum light is caused by
water molecules in the outer region of the atmosphere. 
The corresponding visibility profiles are shown in Fig.~\ref{vis}. 
For each phase, the values of the $K$-band angular radii ($R^{\rm a}_K$) and
the Rosseland angular radii ($R^{\rm a}_{\rm Ross}$) are given.
The arrows mark the position of the Rosseland angular radius.
The protrusion seen for phases $\Phi\,\ge\,0.18$ results from the strong
increase of water absorption behind a shock front caused by the pulsation
of the stellar atmosphere}
  \label{CLV}
\end{figure*}

\begin{table*}
\begin{center}
\caption {Diameter-phase relationship: the columns give (from left to right): 
phase $\Phi$, models used for fitting, Rosseland angular diameter $d_{\rm{Ross}}^{\rm{a}}$ 
(derived by fitting P~model visibility curves from TLSW and ISW to the observed visibilities), 
Rosseland linear radius $R_{\rm Ross}$ derived from the observations, 
theoretical P~model Rosseland radius $R_{\rm{Ross, model}}$, and UD angular diameter.}
\label{diamrad}
\begin{tabular}{c c c c c c c}\hline
$\Phi$ & model & $d_{\rm{Ross}}^{\rm{a}}$ [mas] & $R_{\rm Ross}$ [$R_{\odot}$]
  & $R_{\rm{Ross, model}}$ 
[$R_{\odot}$] & $d_{\rm{UD}}^{\rm{a}}$ [mas]\\ \hline
0.13 & P21n & $28.90 \pm 0.29$ & $332 \pm 38$ & 296 & $29.24 \pm 0.30$ \\ 
0.18 & P22  & $30.82 \pm 0.31$ & $355 \pm 41$ & 318 & $29.53 \pm 0.30$ \\ 
0.26 & P23n & $32.18 \pm 0.32$ & $370 \pm 42$ & 328 & $30.49 \pm 0.30$ \\
0.40 & P24n & $34.92 \pm 0.35$ & $402 \pm 46$ & 333 & $33.27 \pm 0.33$ \\ \hline
\end{tabular}
\end{center}
\end{table*}
Taking the Rosseland diameters derived with the P~model CLVs at the four different phases within the first cycle,  
we obtained the diameter-phase relationship presented in Fig.~\ref{RossRad}.
We do not consider the data within the second cycle of the observations, since the visibilities at $\Phi = 1.40$ 
and $\Phi = 1.47$ lie in a completely different spatial frequency range (or
baseline range, respectively) compared to the data for $\Phi<1$, and 
the derived diameters are very sensitive to small
differences between the assumed model CLV shape and the actual CLV shape of the object.
Figure \ref{RossRad} clearly shows a monotonic increase of the Rosseland angular diameter 
 with decreasing brightness from phase $\Phi = 0.13$ ($d_{\rm{Ross}}^{\rm{a}} = 28.9 \pm 0.3$~mas, 
corresponding to a Rosseland linear radius of $R_{\rm{Ross}} = 332 \pm 38$~$R_{\odot}$) 
to phase $\Phi = 0.40$ ($d_{\rm{Ross}}^{\rm{a}} = 34.9 \pm 0.4$~mas, $R_{\rm{Ross}} = 402 \pm 46$~$R_{\odot}$). 
Therefore, as the visual intensity decreases, the $K$-band diameter of $o$~Cet increases. 
From our analysis we find an 18\% increase of the diameters of $o$~Cet between $\Phi = 0.13$ and $\Phi = 0.40$. 
\cite{THOM} found a similar diameter increase for the oxygen-rich Mira star S~Lac. 
They obtained an increase of the angular size of 22\% between maximum ($\Phi = 0.0$) and minimum light ($\Phi=0.5$). 
This corresponds to a variation of approximately 14\% between $\Phi = 0.13$ and $\Phi = 0.40$ (linear interpolation), 
which is close to our result for $o$~Cet. \\
Our results of the diameter-phase variation are summarized in Table
\ref{diamrad}. For comparison, UD angular diameters were also included. 
As can be seen, $d_{\rm{Ross}}^{\rm{a}}$ is similar to $d_{\rm{UD}}^{\rm{a}}$, 
but the P models agree much better with the measured visibility shapes than
the simple UD models, as we discussed in the previous section. \\
The intensity profiles of the best fitting models for the four different phases,
which were found to be the P21n, P22, P23n, and P24n 
series models, are shown in Fig.~\ref{CLV}. 
The P~models are among those with the lowest effective temperatures and exhibit extended faint wings near maximum light 
($\Phi=0.0$) and protrusions near minimum light ($\Phi=0.5$). 
The extended wing in the intensity profile at maximum light is mainly caused by water molecules, 
which are one of the most abundant molecules in the atmospheres of oxygen-rich Miras (\citealt{TEJ}).
From the analysis of the ISO spectrum, \cite{YAM} concluded that H$_{2}$O layers are more extended at maximum light. 
This is in agreement with the CLV near maximum light (top left panel in Fig.~\ref{CLV}). 
As the visual brightness 
decreases from maximum to minimum phase, the effective temperature decreases, and the extended wings are less pronounced. 
From $\Phi = 0.18$ to $\Phi = 0.40$, water 
absorption increases strongly behind the shock front. 
This results (i) in a
protrusion of the CLV shape 
as seen in R~Leo by \cite{PER} (``Gaussian-type'' CLV) 
and (ii) in substantial molecular-band blanketing of continuum windows and an increase
of $R^{\rm a}_{\rm{Ross}}$,whose position is marked by arrows in Fig.~\ref{CLV} (see the discussion in \citealt{TEJ}). 
Note that the theoretical $R_{\rm Ross}^{\rm a}$, 
as well as the theoretical $R_{1.04}^{\rm a}$ that approximately describes ``continuum pulsation'' 
(Table~\ref{models2}), reaches their maximum values at $\Phi\sim 0.4$ and $\Phi\sim 0.2$, 
respectively (Fig.~\ref{RossRad}), where the difference is caused by the
increasing blanketing of near-continuum windows towards the minimum phase.
Due to the lack of observations close to minimum and maximum light phases and the size of the error bars of the radii, 
there is not enough evidence from the VINCI measurements to confirm this
behavior of $R_{\rm Ross}^{\rm a}$ and $R_{1.04}^{\rm a}$.
Note in this context that the absolute zero-phase point is
uncertain by about 0.05 to 0.1 due to the irregularities of both the observed 
(e.g., \citealt{WHI}) and the model-predicted (HSW, ISW) light curves.\\
The CLVs predicted by the M models have comparable protrusions, and although
the M model visibility fits to the observed visibilities
do not show good agreement, they yield Rosseland linear radii which agree with
the theoretical Rosseland linear radii.
Because of higher temperatures and lower densities, the D models do not
exhibit the protrusions seen in the P and M models. 
Therefore, the slope of the visibility is not as flat as in the case of P and M~models. 
For a more detailed description of the different model CLVs, we refer to BSW, HSW, TLSW, and \cite{TEJ}.
%
\section{Conclusions}\label{concl}

We presented \textit{K}-band observations of $o$~Cet obtained with the VLTI and its beam combiner instrument VINCI.
From these VINCI observations at six different
epochs we derived Rosseland angular radii using 5 different theoretical model series from BSW, HSW, TLSW, and ISW.
Using the derived Rosseland angular diameter and the SEDs reconstructed from various photometry and spectrophotometry data, 
we obtained effective temperatures ranging from $T_{\rm eff}=3192 \pm 200$~K at $\Phi=0.13$ to $2918 \pm 183$~K for $\Phi=0.26$.\\
We found that there is fair agreement between the Rosseland linear radii derived from the observed visibilities and 
those predicted by the fundamental  mode pulsation model series P and M, while there is no agreement for other models.
The nonlinear pulsation models all start from a static ``parent'' star.  
The parent star for the best-fitting model series, the fundamental-mode P series, has a radius of $\sim$240 R$_{\odot}$.  
It is clear from Fig. 4 that $o$~Cet, and indeed the nonlinear
pulsation models from the P series, nearly all have radii of $\sim$300-400 R$_{\odot}$.  
Thus, the effect of large amplitude pulsation is to expand the
surface layers of the star so that its apparent radius is considerably larger at most
phases than the radius it would have if static (note that this effect is milder with respect to a 
continuum radius like $R_{1.04}$ ; cf. Table \ref{models2} and the discussion in ISW).  
This expansion of the apparent radius does not greatly affect the interior of the model or the pulsation period.
Hence, if one compares observed radii of Miras in (say) a radius-period diagram
with the radii and pulsation periods of the parent stars, then one will clearly not find agreement.  
This is the reason that the Miras were thought for so long
to be first-overtone pulsators (see, e.g., \citealt{FEREV,TUCH}, and references therein).  
It is only by a detailed comparison of interferometric
angular diameter measurements with models of large-amplitude, pulsating
atmospheres that this problem has been solved.\\
On the other hand, the effective temperatures derived from the observations are very close to the effective temperatures 
predicted by the D-model series but higher than those predicted by the P and M~models.
Given the definition of effective temperature 
$L = 4 \pi\sigma R^{2} T^{4}_{\rm eff}$, this tells us that the P-series models
are too low in luminosity, consistent with the luminosities derived in section \ref{teff}
for $o$ Cet (compared to the model values in Table \ref{models2}).
The shape of the measured visibilities for $o$~Cet at phase $\Phi=0.13$ are best fitted with the P~model series, whereas 
all other model series and simple UD models show much poorer agreement with the observations.
Taking all this into account, it is clear that a higher-luminosity, fundamental-mode model series is required
for a more accurate modeling of $o$ Cet.\\
Furthermore, we found that the observed visibility functions and diameters change considerably from phase 0.13 to phase 0.40. 
The Rosseland angular diameter of $o$~Cet increases from $d_{\rm{Ross}}^{\rm{a}} = 28.9 \pm 0.3$~mas
(corresponding to a Rosseland linear radius of $R_{\rm Ross} = 332 \pm 38 ~R_{\odot}$) 
at $\Phi=0.13$ to $d_{\rm{Ross}}^{\rm{a}} = 34.9 \pm 0.4$~mas ($R_{\rm Ross} = 402 \pm 46 ~R_{\odot}$) at $\Phi=0.4$. 
Thus, the diameter of $o$~Cet increases by 18\% between these two phases, which
is in good agreement with the approximately 14\% diameter increase derived from linear interpolation of 
the results by \cite{THOM} for the oxygen-rich Mira S~Lac.
\begin{acknowledgements}
We acknowledge with thanks the variable star observations from the AAVSO International 
Database contributed by observers worldwide and used in this research.
This research was in part supported (M.Scholz, P.R.Wood) by the Australian Research Council and
the Deutsche Forschungsgemeinschaft within the linkage project "Red Giants".
We would like to thank Prof. Robert Wing for providing his spectrophotometry measurements.
In addition, we would like to thank Dr. Boris Yudin for his valuable $JHK$ photometry data.
These interferometric measurements have been obtained using the Very Large
Telescope Interferometer, operated by the European Southern Observatory at
Cerro Paranal, Chile.
We thank the VLTI Team (Garching and Paranal) for its support in
this research. All the VINCI data presented in this paper are public and
were retrieved from the ESO/ST-ECF Archive (Garching, Germany).
Finally, we thank the referee for his many useful comments which helped to
improve the manuscript.

\end{acknowledgements}
\bibliographystyle{aa}
\bibliography{0826}

\begin{thebibliography}{62}
\expandafter\ifx\csname natexlab\endcsname\relax\def\natexlab#1{#1}\fi

\bibitem[{{B{\" o}hm-Vitense}(1958)}]{BOV}
{B{\" o}hm-Vitense}, E. 1958, Zeitschrift f\"ur Astrophysik, 46, 108

\bibitem[{{Barnes}(1973)}]{BAR}
{Barnes}, T.~G. 1973, \apjs, 25, 369

\bibitem[{{Baschek} {et~al.}(1991){Baschek}, {Scholz}, \& {Wehrse}}]{BAS}
{Baschek}, B., {Scholz}, M., \& {Wehrse}, R. 1991, \aap, 246, 374

\bibitem[{{Bessell} {et~al.}(1989){Bessell}, {Brett}, {Wood}, \&
  {Scholz}}]{BES}
{Bessell}, M.~S., {Brett}, J.~M., {Wood}, P.~R., \& {Scholz}, M. 1989, \aap,
  213, 209

\bibitem[{{Bessell} {et~al.}(1996){Bessell}, {Scholz}, \& {Wood}}]{BSW}
{Bessell}, M.~S., {Scholz}, M., \& {Wood}, P.~R. 1996, \aap, 307, 481

\bibitem[{{Bonneau} {et~al.}(1982){Bonneau}, {Foy}, {Blazit}, \&
  {Labeyrie}}]{BON82}
{Bonneau}, D., {Foy}, R., {Blazit}, A., \& {Labeyrie}, A. 1982, \aap, 106, 235

\bibitem[{{Bonneau} \& {Labeyrie}(1973)}]{BON}
{Bonneau}, D. \& {Labeyrie}, A. 1973, \apjl, 181, L1

\bibitem[{{Burns} {et~al.}(1998){Burns}, {Baldwin}, {Boysen}, {Haniff},
  {Lawson}, {Mackay}, {Rogers}, {Scott}, {St.-Jacques}, {Warner}, {Wilson}, \&
  {Young}}]{BURNS}
{Burns}, D., {Baldwin}, J.~E., {Boysen}, R.~C., {et~al.} 1998, \mnras, 297, 462

\bibitem[{{Celis S.}(1982)}]{CEL82}
{Celis S.}, L. 1982, \aj, 87, 1791

\bibitem[{{Coud{\' e} du Foresto} {et~al.}(1997){Coud{\' e} du Foresto},
  {Ridgway}, \& {Mariotti}}]{FOR}
{Coud{\' e} du Foresto}, V., {Ridgway}, S., \& {Mariotti}, J.-M. 1997, \aaps,
  121, 379

\bibitem[{{Coud{\' e} du Foresto} \& {Ridgway}(1992)}]{CR}
{Coud{\' e} du Foresto}, V. \& {Ridgway}, S.~T. 1992, in High-Resolution
  Imaging by Interferometry II, eds. Merkle F., Boeckers J.~M., ESO Proc., 731

\bibitem[{{Danchi} {et~al.}(1994){Danchi}, {Bester}, {Degiacomi}, {Greenhill},
  \& {Townes}}]{DAN}
{Danchi}, W.~C., {Bester}, M., {Degiacomi}, C.~G., {Greenhill}, L.~J., \&
  {Townes}, C.~H. 1994, \aj, 107, 1469

\bibitem[{{ESA}(1997)}]{ESA}
{ESA}. 1997, VizieR Online Data Catalog, 1239, 1

\bibitem[{{Feast}(1999)}]{FEREV}
{Feast}, M. 1999, in IAU Symp. 191: Asymptotic Giant Branch Stars, eds.~T.~Le
  Bertre, A.~L\`ebre and C.~Waelkens, 109

\bibitem[{{Feast}(1996)}]{Feast}
{Feast}, M.~W. 1996, \mnras, 278, 11

\bibitem[{{Haniff} {et~al.}(1992){Haniff}, {Ghez}, {Gorham}, {Kulkarni},
  {Matthews}, \& {Neugebauer}}]{HAN92}
{Haniff}, C.~A., {Ghez}, A.~M., {Gorham}, P.~W., {et~al.} 1992, \aj, 103, 1662

\bibitem[{{Haniff} {et~al.}(1995){Haniff}, {Scholz}, \& {Tuthill}}]{HAN95}
{Haniff}, C.~A., {Scholz}, M., \& {Tuthill}, P.~G. 1995, \mnras, 276, 640

\bibitem[{{Hofmann} {et~al.}(2000{\natexlab{a}}){Hofmann}, {Balega}, {Scholz},
  \& {Weigelt}}]{HOFRCAS}
{Hofmann}, K.-H., {Balega}, Y., {Scholz}, M., \& {Weigelt}, G.
  2000{\natexlab{a}}, \aap, 353, 1016

\bibitem[{{Hofmann} {et~al.}(2001){Hofmann}, {Balega}, {Scholz}, \&
  {Weigelt}}]{HOF01}
---. 2001, \aap, 376, 518

\bibitem[{{Hofmann} {et~al.}(2000{\natexlab{b}}){Hofmann}, {Beckmann}, {Bl{\"
  o}cker}, {Coud{\' e} du Foresto}, {Lacasse}, {Millan-Gabet}, {Morel}, {Pras},
  {Ruilier}, {Schertl}, {Scholz}, {Shenavrin}, {Traub}, {Weigelt},
  {Wittkowski}, \& {Yudin}}]{HOF00}
{Hofmann}, K.-H., {Beckmann}, U., {Bl{\" o}cker}, T., {et~al.}
  2000{\natexlab{b}}, in Interferometry in Optical Astronomy, eds.~P.~J.~L\'ena
  and A.~Quirrenbach, SPIE Proc., 4006, 688

\bibitem[{{Hofmann} {et~al.}(1998){Hofmann}, {Scholz}, \& {Wood}}]{HSW}
{Hofmann}, K.-H., {Scholz}, M., \& {Wood}, P.~R. 1998, \aap, 339, 846

\bibitem[{{Ireland} {et~al.}(2004){Ireland}, {Scholz}, \& {Wood}}]{ISW}
{Ireland}, M.~J., {Scholz}, M., \& {Wood}, P.~R. 2004, {MNRAS}, in press

\bibitem[{{Jacob} \& {Scholz}(2002)}]{JS}
{Jacob}, A.~P. \& {Scholz}, M. 2002, \mnras, 336, 1377

\bibitem[{{Karovska} {et~al.}(1991){Karovska}, {Nisenson}, {Papaliolios}, \&
  {Boyle}}]{KAR}
{Karovska}, M., {Nisenson}, P., {Papaliolios}, C., \& {Boyle}, R.~P. 1991,
  \apjl, 374, L51

\bibitem[{{Kervella} {et~al.}(2003){Kervella}, {Gitton}, {Segransan}, {di
  Folco}, {Kern}, {Kiekebusch}, {Duc}, {Longinotti}, {Coud{\' e} du Foresto},
  {Ballester}, {Sabet}, {Cotton}, {Schoeller}, \& {Wilhelm}}]{KER}
{Kervella}, P., {Gitton}, P.~B., {Segransan}, D., {et~al.} 2003, in
  Interferometry for Optical Astronomy II. ed. W.~A.~Traub, SPIE Proc., 4838,
  858

\bibitem[{{Knapp} {et~al.}(2003){Knapp}, {Pourbaix}, {Platais}, \&
  {Jorissen}}]{KNAPP}
{Knapp}, G.~R., {Pourbaix}, D., {Platais}, I., \& {Jorissen}, A. 2003, \aap,
  403, 993

\bibitem[{{Labeyrie}(1970)}]{LAB70}
{Labeyrie}, A. 1970, \aap, 6, 85

\bibitem[{{Labeyrie} {et~al.}(1977){Labeyrie}, {Koechlin}, {Bonneau}, {Blazit},
  \& {Foy}}]{LAB77}
{Labeyrie}, A., {Koechlin}, L., {Bonneau}, D., {Blazit}, A., \& {Foy}, R. 1977,
  \apjl, 218, L75

\bibitem[{{Lockwood} \& {Wing}(1971)}]{LOCK71}
{Lockwood}, G.~W. \& {Wing}, R.~F. 1971, \apj, 169, 63

\bibitem[{{Mariotti} {et~al.}(1996){Mariotti}, {Coud{\' e} du Foresto},
  {Perrin}, {Zhao}, \& {L\'ena}}]{MAR}
{Mariotti}, J.-M., {Coud{\' e} du Foresto}, V., {Perrin}, G., {Zhao}, P., \&
  {L\'ena}, P. 1996, \aaps, 116, 381

\bibitem[{{Mattei}(2003)}]{AAVSO}
{Mattei}, J.~A. 2003, private communication

\bibitem[{{Meisner}(2003)}]{MEIS}
{Meisner}, J.~A. 2003, \apss, 286, 119

\bibitem[{Pease(1931)}]{PEA}
Pease, F.~G. 1931, Ergebn.~d.~Exakten Naturwiss., 10, 84

\bibitem[{{Perrin}(2003)}]{PER2}
{Perrin}, G. 2003, \aap, 400, 1173

\bibitem[{{Perrin} {et~al.}(1999){Perrin}, {Coud{\' e} du Foresto}, {Ridgway},
  {Mennesson}, {Ruilier}, {Mariotti}, {Traub}, \& {Lacasse}}]{PER}
{Perrin}, G., {Coud{\' e} du Foresto}, V., {Ridgway}, S.~T., {et~al.} 1999,
  \aap, 345, 221

\bibitem[{{Quirrenbach} {et~al.}(1992){Quirrenbach}, {Mozurkewich},
  {Armstrong}, {Johnston}, {Colavita}, \& {Shao}}]{QUI}
{Quirrenbach}, A., {Mozurkewich}, D., {Armstrong}, J.~T., {et~al.} 1992, \aap,
  259, L19

\bibitem[{{Richichi} \& {Percheron}(2002)}]{CHARM}
{Richichi}, A. \& {Percheron}, I. 2002, \aap, 386, 492

\bibitem[{{Robertson} \& {Feast}(1981)}]{ROBF}
{Robertson}, B.~S.~C. \& {Feast}, M.~W. 1981, \mnras, 196, 111

\bibitem[{{Scholz}(1985)}]{SCHO}
{Scholz}, M. 1985, \aap, 145, 251

\bibitem[{{Scholz}(2003)}]{SCH03}
{Scholz}, M. 2003, in Interferometry for Optical Astronomy II, ed.~W.~A.~Traub,
  SPIE Proc., 4838, 163

\bibitem[{{Scholz} \& {Takeda}(1987)}]{ST}
{Scholz}, M. \& {Takeda}, Y. 1987, \aap, 186, 200

\bibitem[{{Schwarzschild}(1975)}]{SCHW}
{Schwarzschild}, M. 1975, \apj, 195, 137

\bibitem[{{Smith}(2003)}]{SMITH03}
{Smith}, B.~J. 2003, \aj, 126, 935

\bibitem[{{Smith} {et~al.}(2002){Smith}, {Leisawitz}, {Castelaz}, \&
  {Luttermoser}}]{SMITH02}
{Smith}, B.~J., {Leisawitz}, D., {Castelaz}, M.~W., \& {Luttermoser}, D. 2002,
  \aj, 123, 948

\bibitem[{{Tej} {et~al.}(2003{\natexlab{a}}){Tej}, {Lan{\c c}on}, \&
  {Scholz}}]{TEJ}
{Tej}, A., {Lan{\c c}on}, A., \& {Scholz}, M. 2003{\natexlab{a}}, \aap, 401,
  347

\bibitem[{{Tej} {et~al.}(2003{\natexlab{b}}){Tej}, {Lan\c con}, {Scholz}, \&
  {Wood}}]{TLSW}
{Tej}, A., {Lan\c con}, A., {Scholz}, M., \& {Wood}, P.~R. 2003{\natexlab{b}},
  \aap, 412, 481

\bibitem[{{Thompson} {et~al.}(2002){Thompson}, {Creech-Eakman}, \& {van
  Belle}}]{THOM}
{Thompson}, R.~R., {Creech-Eakman}, M.~J., \& {van Belle}, G.~T. 2002, \apj,
  577, 447

\bibitem[{{Tuchman}(1999)}]{TUCH}
{Tuchman}, Y. 1999, in IAU Symp. 191: Asymptotic Giant Branch Stars, eds.~T.~
  Le Bertre, A.~L\`ebre and C.~Waelkens, 123

\bibitem[{{Tuthill} {et~al.}(1994){Tuthill}, {Haniff}, \& {Baldwin}}]{TUT94}
{Tuthill}, P.~G., {Haniff}, C.~A., \& {Baldwin}, J.~E. 1994, in IAU Symp. 158:
  Very High Angular Resolution Imaging, eds. J.~G. Robertson and J.~T.~William,
  395

\bibitem[{{van Belle} {et~al.}(1996){van Belle}, {Dyck}, {Benson}, \&
  {Lacasse}}]{VANB}
{van Belle}, G.~T., {Dyck}, H.~M., {Benson}, J.~A., \& {Lacasse}, M.~G. 1996,
  \aj, 112, 2147

\bibitem[{{Watanabe} \& {Kodaira}(1979)}]{WAN}
{Watanabe}, T. \& {Kodaira}, K. 1979, \pasj, 31, 61

\bibitem[{{Weigelt} {et~al.}(1996){Weigelt}, {Balega}, {Hofmann}, \&
  {Scholz}}]{WEI96}
{Weigelt}, G., {Balega}, Y., {Hofmann}, K.-H., \& {Scholz}, M. 1996, \aap, 316,
  L21

\bibitem[{{Weigelt} {et~al.}(2003){Weigelt}, {Beckmann}, {Berger}, {Bloecker},
  {Brewer}, {Hofmann}, {Lacasse}, {Malanushenko}, {Millan-Gabet}, {Monnier},
  {Ohnaka}, {Pedretti}, {Schertl}, {Schloerb}, {Scholz}, {Traub}, \&
  {Yudin}}]{WEI03}
{Weigelt}, G., {Beckmann}, U., {Berger}, J., {et~al.} 2003, in Interferometry
  for Optical Astronomy II. ed. W.~A.~Traub, SPIE Proc., 4838, 181

\bibitem[{{Weigelt} {et~al.}(2000){Weigelt}, {Mourard}, {Abe}, {Beckmann},
  {Chesneau}, {Hillemanns}, {Hofmann}, {Ragland}, {Schertl}, {Scholz}, {Stee},
  {Thureau}, \& {Vakili}}]{WEI00}
{Weigelt}, G., {Mourard}, D., {Abe}, L., {et~al.} 2000, in Interferometry in
  Optical Astronomy, eds.~P.~J.~L\'ena and A.~Quirrenbach, SPIE Proc., 4006,
  617

\bibitem[{{Weiner} {et~al.}(2000){Weiner}, {Danchi}, {Hale}, {McMahon},
  {Townes}, {Monnier}, \& {Tuthill}}]{WEIN}
{Weiner}, J., {Danchi}, W.~C., {Hale}, D.~D.~S., {et~al.} 2000, \apj, 544, 1097

\bibitem[{{Whitelock} {et~al.}(2000){Whitelock}, {Marang}, \& {Feast}}]{WHI}
{Whitelock}, P., {Marang}, F., \& {Feast}, M. 2000, \mnras, 319, 728

\bibitem[{{Wilson} {et~al.}(1992){Wilson}, {Baldwin}, {Buscher}, \&
  {Warner}}]{WIL92}
{Wilson}, R.~W., {Baldwin}, J.~E., {Buscher}, D.~F., \& {Warner}, P.~J. 1992,
  \mnras, 257, 369

\bibitem[{{Wing}(2003)}]{RW}
{Wing}, R. 2003, private communication

\bibitem[{{Wittkowski} {et~al.}(2004){Wittkowski}, {Aufdenberg}, \&
  {Kervella}}]{psiphe}
{Wittkowski}, M., {Aufdenberg}, J.~P., \& {Kervella}, P. 2004, \aap, 413, 711

\bibitem[{{Wood} {et~al.}(1999){Wood}, {Alcock}, {Allsman}, {Alves}, {Axelrod},
  {Becker}, {Bennett}, {Cook}, {Drake}, {Freeman}, {Griest}, {King}, {Lehner},
  {Marshall}, {Minniti}, {Peterson}, {Pratt}, {Quinn}, {Stubbs}, {Sutherland},
  {Tomaney}, {Vandehei}, \& {Welch}}]{woodM}
{Wood}, P.~R., {Alcock}, C., {Allsman}, R.~A., {et~al.} 1999, in IAU Symp. 191:
  Asymptotic Giant Branch Stars, eds.~T.~ Le Bertre, A.~L\`ebre and
  C.~Waelkens, 151

\bibitem[{{Yamamura} {et~al.}(1999){Yamamura}, {de Jong}, \& {Cami}}]{YAM}
{Yamamura}, I., {de Jong}, T., \& {Cami}, J. 1999, \aap, 348, L55

\bibitem[{{Yudin}(2003)}]{BY}
{Yudin}, B. 2003, private communication

\end{thebibliography}
\end{document}